# Analysis of Fe and Co binary catalysts in chemical vapor deposition growth of single-walled carbon nanotubes

Qingmei Hu[1], Ya Feng[1,2*], Wanyu Dai[1], Daisuke Asa[1], Daniel Hedman[3], Aina Fitó Parera[4], Yixi Yao[5], Yongjia Zheng[1,6], Kaoru Hisama[7], Gunjan Auti[1], Hirofumi Daiguji[1], Christophe Bichara[8], Shohei Chiashi[1], Yan Li[5], Wim Wenseleers[9], Dmitry Levshov[4], Sofie Cambré[4], Keigo Otsuka[1], Rong Xiang[1,6], Shigeo Maruyama[1,6*]

[1]Department of Mechanical Engineering, The University of Tokyo, Tokyo, 113-8656, Japan

[2]Key Laboratory of Ocean Energy Utilization and Energy Conservation of Ministry of Education, School of Energy and Power Engineering, Dalian University of Technology, Dalian, Liaoning 116024, China

[3]Center for Multidimensional Carbon Materials (CMCM), Institute for Basic Science (IBS), Ulsan, 44919, Republic of Korea

[4]Theory and Spectroscopy of Molecules and Materials, Department of Physics and Department of Chemistry, University of Antwerp, Antwerp 2610, Belgium

[5]Center Beijing National Laboratory for Molecular Science, Key Laboratory for the Physics and Chemistry of Nanodevices, State Key Laboratory of Rare Earth Materials Chemistry and Applications, College of Chemistry and Molecular Engineering, Peking University, Beijing, 100871, China

*[6]State Key Laboratory of Fluid Power and Mechatronic Systems, School of Mechanical Engineering, Zhejiang University, Hangzhou 310027, China*

*[7]Interdisciplinary Cluster for Cutting Edge Research, Research Initiative for Supra-Materials, Shinshu University, Japan*

*[8]The Interdisciplinary Nanoscience Centre of Marseille (CINaM), Aix-Marseille University and CNRS, Marseille, France*

*[9]Nanostructured and Organic Optical and Electronic Materials, Department of Physics, University of Antwerp, Antwerp 2610, Belgium*

\* Corresponding authors: Shigeo Maruyama, Ya Feng

E-mail addresses: maruyama@photon.t.u-tokyo.ac.jp; fengya@dlut.edu.cn

**ABSTRACT**: Metal catalysts play a pivotal role in the growth of single-walled carbon nanotubes (SWCNTs), with binary metallic catalysts emerging as an efficient SWCNT synthesis strategy. Among these, iron (Fe), cobalt (Co), and their alloys are particularly effective. However, prior studies have predominantly employed Fe-Co alloy catalysts with fixed atomic ratios as well as unchanged chemical vapor deposition (CVD) conditions, leaving the influence of variable Fe-Co compositions and CVD growth parameters on SWCNT synthesis poorly understood. This study focuses on the role of Fe-Co catalyst ratios, with the aim of elucidating the distinct contributions of Fe and Co atoms, in the growth of SWCNTs. By systematically exploring a wide range of Fe-Co ratios and growth conditions, we identified $Fe_{0.75}Co_{0.25}$ as a highly efficient binary catalysts at 850 °C, primarily forming catalyst clusters with diameters of 2.5–6 nm and yielding SWCNTs with diameters ranging from 0.9-1.1 nm. On the other hand, $Fe_0Co_1$ exhibited higher catalytic activity at 600 °C, generating smaller catalyst clusters of 1.5–5 nm and producing SWCNTs with reduced diameters about 0.6–0.9 nm. Transmission Electron Microscope (TEM) and Energy Dispersive X-ray Spectroscopy (EDS) analyses reveal that high SWCNT yields correlate with the formation of uniform sized Fe-Co catalyst particles with surface segregated Co that optimize carbon solubility. Molecular dynamics (MD) simulations further corroborate these findings, demonstrating that the structure and melting behavior of $Fe_xCo_{1-x}$ clusters depend on cluster size and composition.

**INTRODUCTION**

Single-walled carbon nanotubes (SWCNTs) have garnered significant attention for their unique one-dimensional structure and exceptional electronic properties, including structure-dependent optical properties and a chirality-dependent conductivity.[1-4] These characteristics position SWCNTs as highly advantageous candidates for a wide range of future applications, particularly in nano-optoelectronics.[5, 6] However, despite notable advancements in device-level performance, the widespread adoptions of SWCNTs in electronic and optoelectronic technologies remain hindered by their intrinsic heterogeneity. This is particularly pronounced in structural inhomogeneity, such as variations in diameter, chirality, and length, *etc.*, arising from limited catalyst selectivity and sensitivity to growth conditions.[7-10] Chemical vapor deposition (CVD) is widely regarded as a highly effective method for synthesizing carbon nanotubes (CNTs), wherein metal catalysts play a pivotal role in nucleation and growth [11-13]. Extensive theoretical[14, 15] and experimental[16-21] studies have emphasized that the structural and physicochemical properties of catalysts, as well as their compatibility to CVD conditions, are crucial determinants for the efficiency and structural control of SWCNT growth.

To achieve high yield, high purity, and structural selectivity in SWCNT synthesis, considerable efforts have been devoted to developing bimetallic catalysts. Prior studies have demonstrated that specific bimetallic systems, such as Fe-Co,[16, 19] Fe-Cu,[20] Fe-Ni,[22] Fe-Ru,[23] Ru-Co,[24] Co-Mo,[18, 25] Co-Cu,[26] Co-W[27, 28], can facilitate high-growth yields and at the same time enable partial control over SWCNT diameter and chiral structures. For instance, early work by Murakami *et. al.*[29] highlighted the advantage of $Fe_{0.5}Co_{0.5}$ bimetallic systems over monometallic Fe or Co catalysts, showing that Fe effectively inhibited the aggregation of nanoscale Co particles, thereby promoting SWCNT formation. Chiang *et. al.*[30] and Xiang *et. al.*[31] further reported that tuning the composition of $Ni_xFe_{1-x}$ and $Co_xMo_{1-x}$ catalysts can modulate the chirality distribution of as-grown SWCNTs. Li *et. al.*[32] elucidated the relationship between nanotube diameter and catalyst particle size under different growth modes, underscoring the critical role of crystalline stability in chirality-selective growth. Previous studies have primarily explored isolated compositional ratios or specific growth conditions, yet a systematic and mechanistic approach to catalyst tuning is still missing.

In this study, using $Fe_xCo_{1-x}$ nanoparticles (NPs) as catalysts, we systematically investigated the influence of atomic composition in Fe-Co bimetallic catalysts on SWCNT growth efficiency and diameter distribution across a broad CVD temperature window. Spectroscopic analysis revealed that at a relatively low growth temperature (600 °C, in the following text, unless otherwise specified; and the ethanol partial pressure is 50 Pa), $Fe_0Co_1$ catalysts resulted in highest SWCNT yield with diameters in the range of 0.7–0.9 nm. The nanotube yield decreased gradually with increasing Fe in the catalysts due to the enhanced carbon deposition and the emergence of multi-walled carbon nanotubes (MWCNTs) and impurities.[22] Concurrently, the catalyst particle sizes increased with more Fe in the catalysts, resulting in a broader and shift to larger diameter SWCNT distribution. In contrast, at a higher temperature (850 °C, similarly, unless otherwise specified; and the ethanol partial pressure is 50 Pa), the $Fe_{0.75}Co_{0.25}$ catalyst exhibited a markedly enhanced SWCNT growth efficiency, significantly outperforming $Fe_1Co_0$, $Fe_0Co_1$, and other $Fe_xCo_{1-x}$. This improvement is attributed to the superior thermal stability of the binary catalyst with a specific proportion at higher temperatures, which facilitates the formation of smaller, more uniform and more stable NPs and thereby promotes efficient SWCNT growth.

Transmission electron microscopy (TEM) and elemental mapping analyses confirm the formation of uniform sized Fe-Co catalyst particles with surface segregated Co. Furthermore, machine learning force field (MLFF) driven molecular dynamics (MD) simulations show that the structure of Fe-Co catalyst particles changes from a BCC Wulff shape to an icosahedral shape with increasing Co content. This Co-dependent structure transition affects both carbon solubility and size distribution of the experimentally obtained clusters.

**RESULT AND DISCUSSION**

To systematically investigate the impact of atomic ratio in bimetallic catalysts on SWCNT growth, we conducted the experiments within a broad operational window of alcohol catalytical chemical vapor deposition (ACCVD), as outlined in the previous study.[33] A range of pressure and temperature combinations were tested, as shown in **Fig. S1a**, to evaluate how the effect of catalyst composition varies under different growth conditions.. The as-grown SWCNT samples were dispersed in solutions, as described in the Characterization section (as well as SI **Fig. S2b**), and their absorbance was measured to assess growth yield. To more accurately evaluate growth efficiency, a baseline subtraction procedure was applied, as illustrated in **Fig. S1b**. Additionally, thermogravimetric analysis (TGA) was performed (**Fig. S3a**). The relative growth yield, determined from absorbance (**Fig. S3b**), showed a very good agreement with the TGA weight loss (**Fig. S3c**), validating the use of absorbance measurements as a reliable indicator of SWCNT yield.

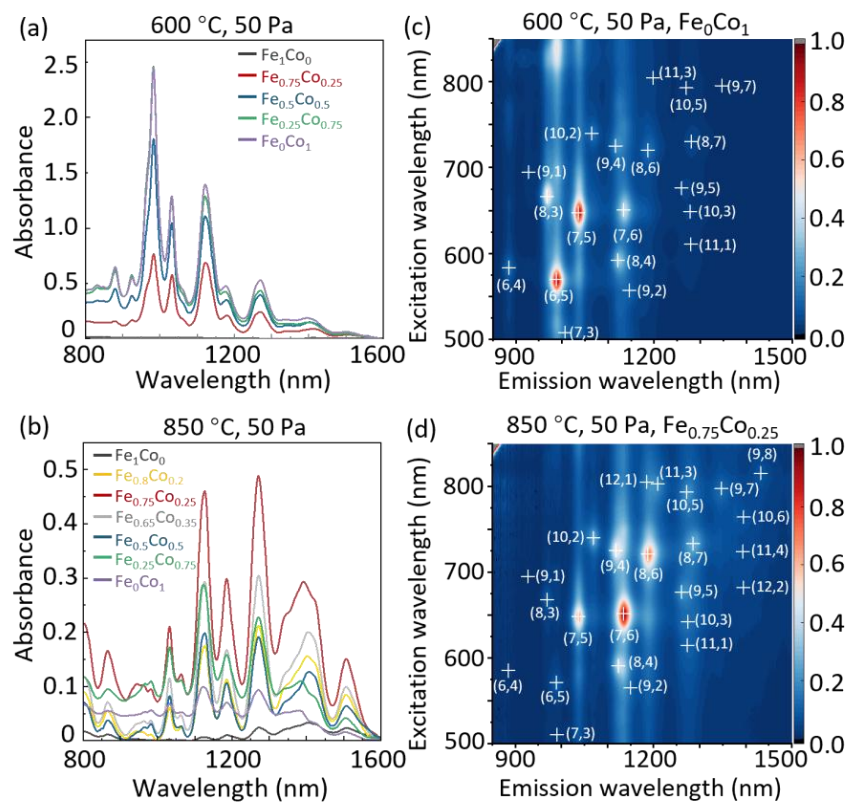

**Figure 1.** Optical characterization. The absorption spectra of dispersed solution samples grown at (a) 600 °C and (b) 850 °C. The PLE maps of samples grow with (c) $Fe_0Co_1$ catalysts, at 600 °C and (d) $Fe_{0.75}Co_{0.25}$ catalysts, at 850 °C.

**Influence of Fe-Co ratio on SWCNT growth**

**Fig. S4–S6** in the Supporting Information (SI) systematically present and analyze the actual catalytic performance of Fe/Co bimetallic catalysts under various growth conditions, from the perspective of total SWCNT yield. The results suggest that under harsher conditions, Fe plays a more significant role in enhancing growth efficiency, whereas under milder or optimized conditions, the catalytic activity is predominantly governed by Co. Given the extremely low yield observed under the 900 °C, 50 Pa condition (The intensity of absorbance are lower than 0.1), which rendered further analysis impractical, two representative growth regimes were selected for subsequent in-depth investigation: (i) a low-temperature, low-pressure regime (600 °C, 50 Pa), and (ii) a high-temperature, low-pressure regime (850 °C, 50 Pa)(**Fig. 1**). Accordingly, the following analyses focus on these two growth windows to investigate the relationship between the roles of Fe and Co and their influence on the yield of SWCNTs with various chirality, the diameter distribution of the SWCNTs, and the corresponding catalyst cluster size. Seven groups with different $Fe_xCo_{1-x}$ catalysts were employed to synthesize SWCNTs under these conditions. The corresponding absorbance spectra were subjected to chirality-resolved fitting to enable further analysis of the yield of individual (n, m) species and the diameter distribution of the SWCNTs.

At 600 °C (**Fig. 1a**), the relative intensity distribution of absorbance peaks for samples grown using different $Fe_xCo_{1-x}$ catalysts was basically consistent. For instance, the peak near 983 nm, primarily attributed to the (6, 5) chirality, was more prominent than other peaks (other species), suggesting that chirality distribution remained relatively stable regardless of the Fe-Co ratio, provided that Co was included in the catalyst composition. Furthermore, the PL peak corresponding to these (6,5) SWCNTs (around 987 nm) was more pronounced (**Fig. 1c** and **S7**), which aligns with the tendency of lower growth temperatures favoring the produce of smaller-diameter SWCNTs.[33-35] It is also worth noting that monometallic Fe exhibited negligible catalytic effect for SWCNT synthesis, whereas monometallic Co achieved the highest absorbance among all $Fe_xCo_{1-x}$ catalysts. This suggests that under the given conditions, Co atoms exhibit intrinsically higher catalytic activity, and the introduction of Fe atoms appears to suppress the catalytic performance of Co.

At 850 °C (**Fig. 1b**), the catalysts preferentially facilitated the growth of large-diameter SWCNTs, while the chirality distribution, except for the range of 1300 to 1450 nm, remained mainly consistent across samples prepared with different $Fe_xCo_{1-x}$ catalysts. Unlike the growth behavior observed at 600 °C, monometallic Fe demonstrated measurable catalytic activity at this elevated temperature, albeit with a modest yield. Among different $Fe_xCo_{1-x}$ catalysts, $Fe_{0.75}Co_{0.25}$, exhibited the highest growth efficiency. Chiral peak intensities in **Fig. S9** were extracted from the normalized PLE spectra (**Fig. S7 and S8**), revealing a distinct shift in distribution, with the $Fe_{0.75}Co_{0.25}$ catalyst at 850 °C favoring larger-diameter species.

To gain deeper insight into controlled growth behavior, an advanced correlated PLE and absorbance fitting algorithm was employed, as shown in **Fig. 2a**, with a detailed comprehensive description provided in the Data processing part and SI **Section 2**. By fitting the absorbance spectra of each $Fe_xCo_{1-x}$ sample individually, the absorbance intensity of different chiral SWCNT species within each sample was quantitatively extracted. **Fig. 2b** and **2d** present representative chiralities with strong absorbance signals under growth temperatures of 600 °C and 850 °C, respectively. At 600 °C, a consistent trend emerged: the absorbance intensity of each chiral species exhibited a positive correlation with the Co content in the catalysts. In contrast, at 850 °C, the sample prepared using the $Fe_{0.75}Co_{0.25}$ catalysts consistently displayed the strongest absorbance peaks, highlighting its superior catalytic efficiency for SWCNT growth. This growth efficiency observed with the $Fe_{0.75}Co_{0.25}$ catalysts suggested that the Fe-Co ratio possibly plays a critical role in modulating the solubility and precipitation dynamics of carbon atoms or the size of the catalyst cluster, thereby significantly influencing the growth efficiency of SWCNTs.

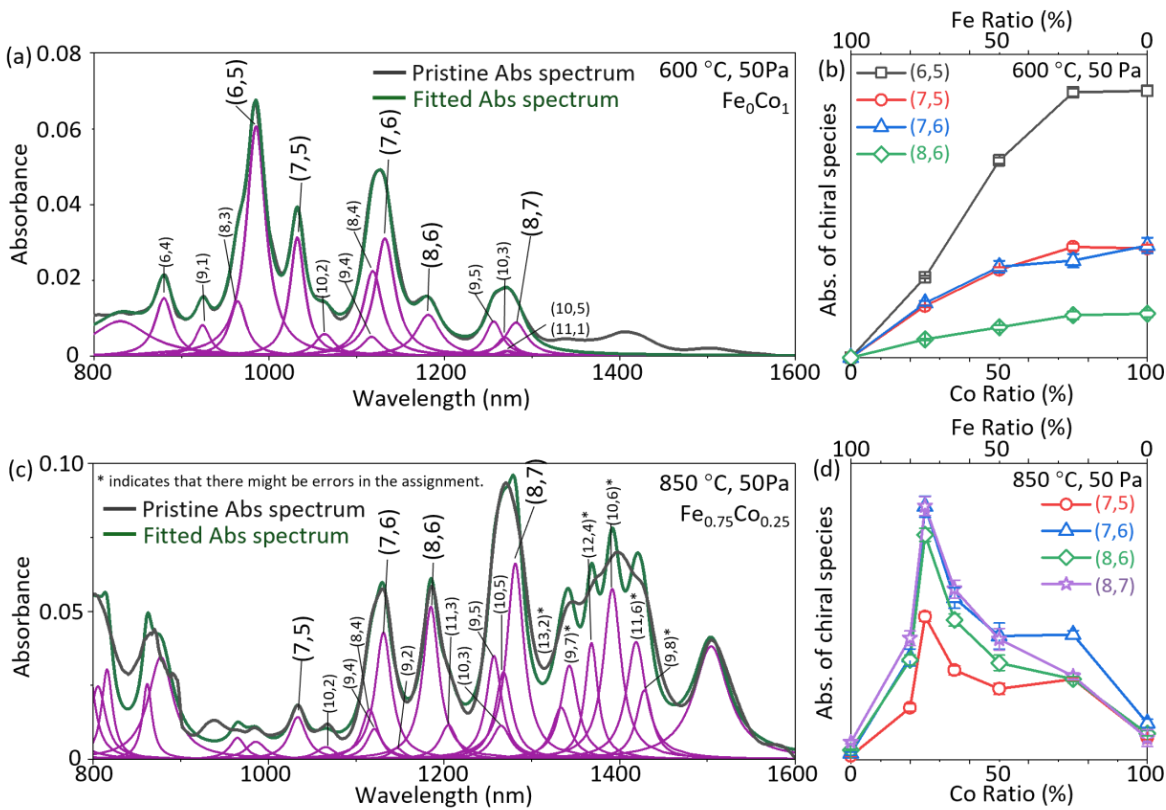

**Figure 2.** Analysis of absorption spectra. (a, c) Fitted absorption spectrum with chiral assignments for the sample prepared using (a) $Fe_0Co_1$ catalysts grown at 600 °C and (c) $Fe_{0.75}Co_{0.25}$ catalyst grown at 850 °C. The absorption feature near 1400 nm involves multiple overlapping chiral peaks. Due to limited intensity of these peaks in the current PLE data, the fitted peak positions in this region may carry significant uncertainty. However, the results remain adequate for the purposes of this study, as all data within one series of experiments is fitted in exactly the same manner and only intensities are left to vary. (b, d) Absorbance dependence of each chiral species on the Co ratio in the catalyst for samples grown at (b) 600 °C and (d) 850 °C.

**TEM characterization and substrate influence on SWCNT growth**

TEM characterizations provide direct insights into the atomic structure of catalyst clusters and as-grown SWCNTs, offering critical insights into the growth process to enhance the growth efficiency in future works. Leveraging recent advancements in imaging techniques,[36, 37] we visualized catalyst particles directly on thin $SiO_2$ films both before and after the SWCNT growth. A key question is whether the substrate material, $SiO_2$/Si or zeolite in this case, affect the final SWCNT products. Specifically, whether the atomic structures observed via TEM correlate with the SWCNT growth yield on zeolite, as evaluated by absorbance measurements. To compare the effects of different substrates on the structural characteristics of SWCNTs and to verify the feasibility of using $SiO_2$/Si as substrates for in-situ growth and subsequent characterization, Raman spectroscope with four excitation lasers (488, 532, 633, and 785 nm) was used to evaluate two sets of samples (#1 synthesized at 800 °C and #2 synthesized at 600 °C) grown on both $SiO_2$/Si and zeolite substrates. As shown in **Fig. 3**, the radial breathing mode (RBM) from both substrates exhibits highly consistent profiles, indicating that the chiral distributions of the SWCNTs are essentially similar, regardless of the substrate type. This suggests that the substrate has minimal influence on chirality distribution under the growth conditions used. It is noteworthy that the RBM features of SWCNTs grown on $SiO_2$/Si substrates exhibit slightly broader peaks under certain excitation conditions, particularly in the low-frequency region or in sample #2, which was synthesized at a relatively low temperature of 600 °C. This broadening likely arises from partial bundling of nanotubes on the $SiO_2$/Si surface or the interaction with the surface, which introduces inhomogeneous resonance conditions and affects peak shapes. However, this difference does not significantly alter the overall chiral distribution, nor does it impede effective spectral analysis. Meanwhile, as shown in **Fig. S8**, the G and D bands are almost resemble across the two substrates, with quite good and consistent G/D ratios, indicating high structural quality. These results demonstrate that the use of $SiO_2$/Si does not significantly alter the growth behavior or product quality, validating their applicability as a dual-purpose substrate for both growth and subsequent SiO$_2$/Si-based structural characterizations.

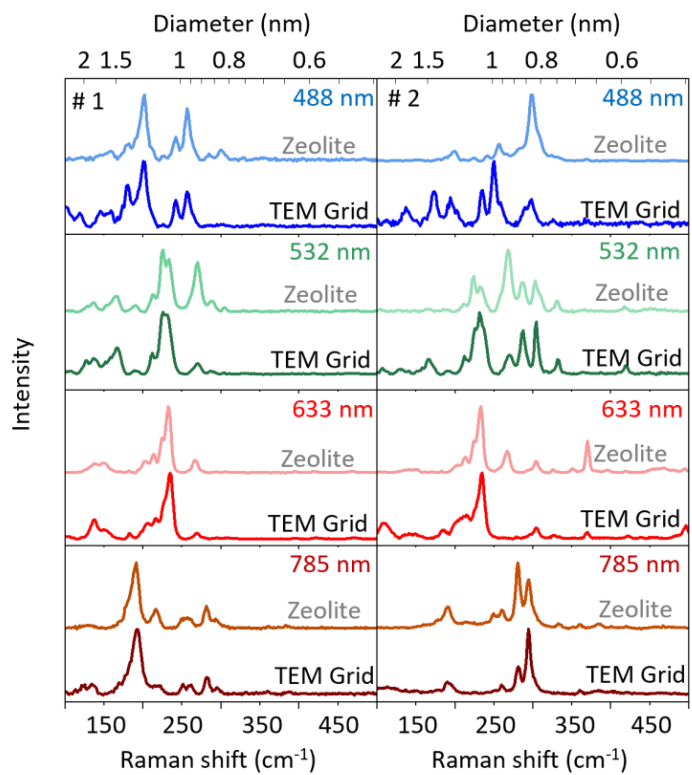

**Figure 3.** Raman spectra (in the RBM range) of two sets of samples (#1 is Fe$_{0.5}$Co$_{0.5}$ and #2 is Fe$_0$Co$_1$, synthesized at 800 °C, 1.2 kPa and 600 °C, 50 Pa, respectively.) grown on zeolite or on a SiO$_2$/Si TEM grid.

Based on the preceding discussions, we prepared varying Fe$_x$Co$_{1-x}$ catalyst samples (Fe$_1$Co$_0$, Fe$_{0.75}$Co$_{0.25}$, Fe$_{0.5}$Co$_{0.5}$, and Fe$_0$Co$_1$) under identical synthesis conditions (850 °C) to those used for the zeolite samples from which absorption spectra were obtained. These catalyst samples were deposited on SiO$_2$/Si TEM grid and individually characterized by TEM to investigate their morphological properties. Subsequently, statistical evaluations were performed to analyze the catalyst NPs. As illustrated in **Fig. 4**, the catalyst NPs were examined both before and after the ACCVD process. This comparative analysis allowed us to assess the atomic structural evolution of the catalyst particles and their influence on SWCNT growth.

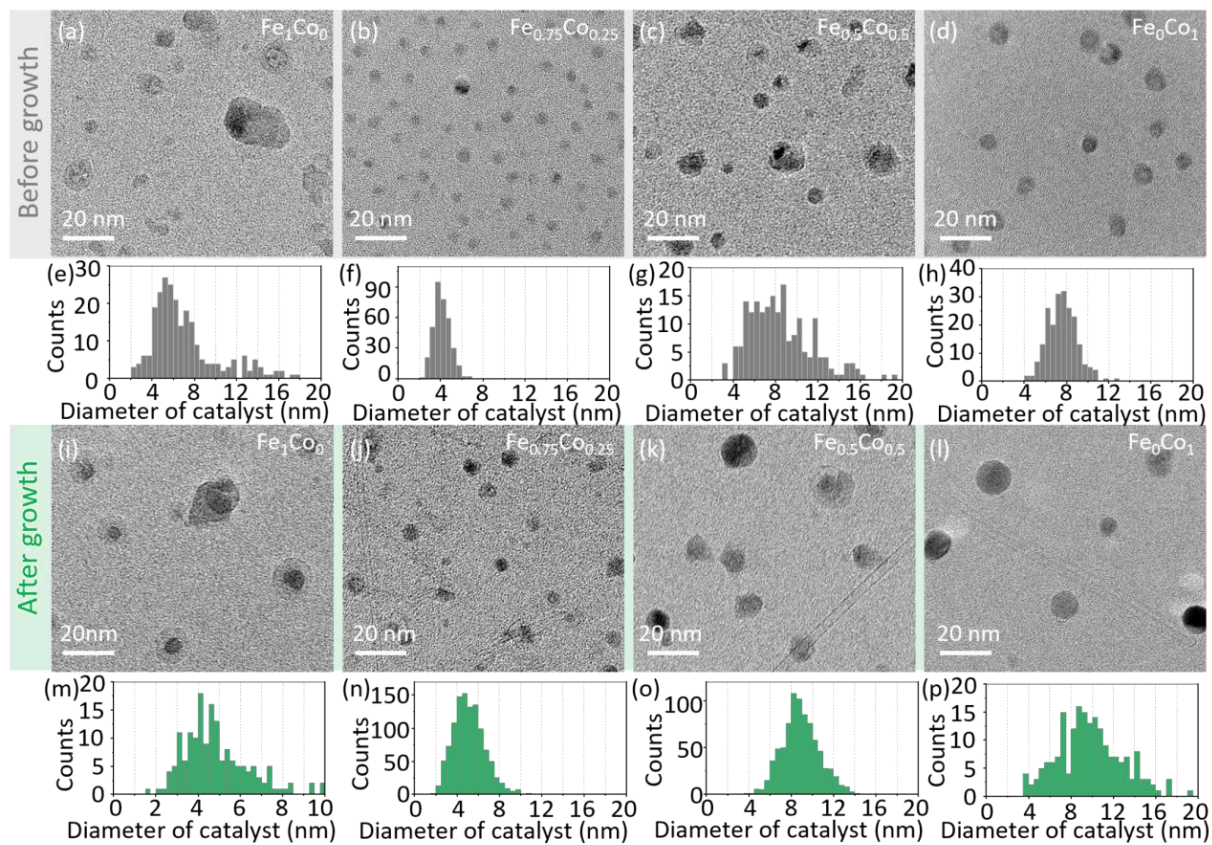

**Figure 4.** TEM characterization and analysis of samples synthesized at 850 °C using different $Fe_xCo_{1-x}$ catalysts. The TEM images of $Fe_xCo_{1-x}$ catalyst particles before (a-d) and after (i-l) SWCNT growth revealing distinct differences in particle size. Corresponding histograms of particle size distributions for catalysts before (e-h) and after (m-p) SWCNT growth.

The morphological structure of catalyst particles before ACCVD is a critical determinant of growth behavior. After reduction in Ar/ $H_2$ gas (3% $H_2$) and prior to growth, the particle size varied significantly depending on the Fe-Co ratio, as illustrated in **Fig. 4a–d**. Specifically, the $Fe_{0.75}Co_{0.25}$ catalyst exhibited a uniform particle distribution and relatively small particle sizes. In contrast, pure Co formed uniform particles of relatively larger sizes. Conversely, the $Fe_1Co_0$ catalyst and $Fe_{0.5}Co_{0.5}$ catalyst displayed irregular and heterogeneous particle size distributions. Statistical analysis further corroborated these observations, revealing that the $Fe_{0.75}Co_{0.25}$ catalyst had the smallest average particle size, ranging from approximately 3 to 4 nm, as can be found in **Fig. 4e–h**. Following the ACCVD process, considerable differences in growth efficiency and particle morphology were observed among the catalysts. The $Fe_{0.75}Co_{0.25}$ catalyst demonstrated the highest growth efficiency, which was attributed to its consistently small particle size throughout the process. Importantly, the particle size of the $Fe_{0.75}Co_{0.25}$ catalyst maintained the smallest even after SWCNT growth, signifying its superior catalytic performance. For the low-yield samples under these conditions—namely, $Fe_0Co_1$ and $Fe_1Co_0$—the catalyst particles were found to be encapsulated by an outer layer of deposited carbon after the ACCVD growth process. This carbon overcoating serves as key evidence of catalyst deactivation, likely caused by an imbalance between carbon dissolution and precipitation dynamics. These results emphasized the critical influence of catalyst composition, and consequently particle size, on both the growth efficiency and of SWCNTs.

Furthermore, we conducted the same statistical analysis on catalyst particles for samples subjected to ACCVD at 600 °C, as presented in **Fig. S9**. In brief, the particle size decreased with increasing Co content prior to ACCVD, leading to higher SWCNT yields after growth, which is consistent with the absorbance trends discussed earlier. A detailed discussion is provided in SI.

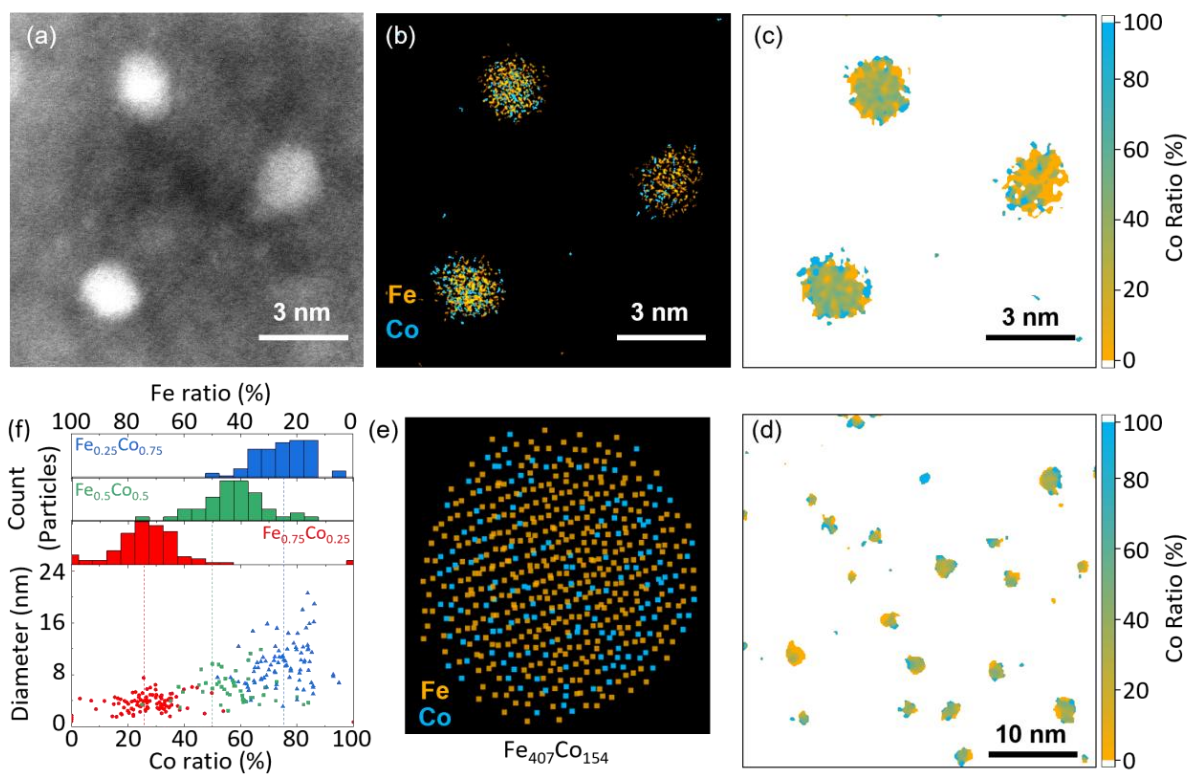

**Figure 5.** Statistical analysis of the Fe–Co ratio in samples grown at 850 °C, obtained by EDS. (a–c) High-magnification (small-area) and (d) low-magnification (large-area) STEM images of $Fe_{0.75}Co_{0.25}$ catalysts after ACCVD growth at 850 °C. Specifically, (a) shows annular dark-field (ADF) images; (b) presents EDS elemental maps of Fe and Co distributions; (c, d) display calculated Co-ratio maps, where the Fe/Co ratio was derived from pixel-by-pixel EDS analysis by measuring the signal intensities of Fe and Co within each cluster. (e) Structure obtained from MD simulations with a slow cooling rate 10 K ns$^{-1}$. (f) Histogram and scatter plot of the experimentally measured Fe–Co ratios in samples grown at 850 °C.

The nominal Co ratio in the catalysts was initially determined by the weight ratio of Co to Fe used during catalyst preparation. However, to confirm the actual $Fe_xCo_{1-x}$ composition in the synthesized samples, elemental analysis was performed across multiple regions. **Fig. 5a** display the corresponding ADF-TEM images of the sample after ACCVD, while **Fig. 5b** present elemental mapping of small areas at high magnifications. The results reveal a high degree of overlap between Co and Fe atoms across all magnifications, indicating that most of the particles are composed of mixed $Fe_xCo_{1-x}$ alloy, confirming the formation of $Fe_xCo_{1-x}$ alloy clusters. Furthermore, the calculated Co ratio mapping, obtained by quantifying the EDS signal intensities of Fe and Co within each cluster and deriving the Fe-Co ratio, is shown in **Fig. 5c** (high magnification) **and 5d** (low magnification). In these maps, yellow regions correspond to high Fe content, while blue regions indicate high Co content. The maps clearly illustrate the actual Fe-Co ratio for individual particles. Notably, most particle cores exhibit colors ranging from deep yellow to deep blue, suggesting the presence of $Fe_xCo_{1-x}$ alloys in the core regions. In contrast, the outer surfaces of these particles are mainly yellow, indicating surface segregated Co regions. This observation points to the formation of alloy nanoparticles with surface segregated Co regions, at an approximate composition of $Fe_{0.75}Co_{0.25}$. A slow cooling MD simulation at 10 K $ns^{-1}$ was performed to obtain the structure of a $Fe_{407}Co_{154}$ (approx. $Fe_{0.75}Co_{0.25}$) nanoparticle. After cooling a BCC Wulff shaped structure was obtained and as seen in **Fig. 5e**, although much smaller than the experimentally observed ones, this nanoparticle shows a similar distribution of Fe and Co atoms.

To further understand the structure of the catalyst particles slow cooling MD simulations at 10 K $ns^{-1}$ was performed to obtain the structure of a $Fe_{407}Co_{154}$ (approx. $Fe_{0.75}Co_{0.25}$) nanoparticle. After cooling a BCC Wulff shaped structure is obtained and as seen in **Fig. 5e**, although much smaller than the experimentally observed ones, this nanoparticle shows a similar distribution of Fe and Co atoms. From the simulation is clear that Co preferentially segregate to the surface of the cluster, which is consistent with the experimental EDS-based Co distribution trends in **Fig. 5c** and **5d**. This strongly supports the formation of thermodynamically stable Fe–Co alloy catalysts under ACCVD growth conditions.

The statistical analysis of elemental mapping data for samples with nominal compositions of $Fe_{0.75}Co_{0.25}$, $Fe_{0.5}Co_{0.5}$, and $Fe_{0.25}Co_{0.75}$ is presented in the histogram and scatter plot in **Fig. 5f.** The results indicate that the actual Fe-Co ratios of the particles closely match the initial weight-based estimations. For example, the $Fe_{0.75}Co_{0.25}$ sample exhibits a Co concentration of approximately 25%, confirming the accuracy of the weight-based Fe-Co ratio estimation during catalyst preparation. Therefore, under the ACCVD condition of 850 °C and 50 Pa, the $Fe_{0.25}Co_{0.75}$ catalyst sample formed particles with an Fe-Co ratio of approximately 25%, achieving efficient utilization of the catalyst without significant loss of atoms with a non-uniform ratio. This efficient catalyst utilization contributed to the highest abundance of SWCNTs and the greatest growth efficiency among all the tested samples, as evidenced by absorption spectra and TEM analysis discussed earlier.

Based on these findings, we hypothesized that NPs with a Fe-Co ratio of approximately 25% and sizes of 2.5-6 nm may strike a favorable balance between thermodynamic stability and catalytic activity. To elucidate the composition and size dependency of the melting point of $Fe_xCo_{1-x}$ clusters several slow cooling followed by slow heating MD simulations where performed. As shown in **Fig. 6** the melting point increases with the size of the cluster and for clusters larger than around 300 atoms the melting point is above 850 °C for all Fe/Co ratios. Since the $Fe_{0.75}Co_{0.25}$ clusters observed in the experiments are significantly larger than this, their melting point is likely higher than the 850 °C reached during ACCVD. For clusters smaller than 300 atoms the melting point increases linearly with the amount of Co reaching its maximum at pure Co clusters. The 561-atom cluster shows an initial decrease in the melting point for increasing Co content reaching its lowest values after around 35% Co. This can be explained by the fact that for larger clusters there exists a Co-dependent phase transition from BCC, most stable for pure Fe clusters, to icosahedral, most stable for pure Co clusters. This competition between BCC and icosahedral structures results in melting point suppression for the 561-atom cluster at Co ratios between ~35 and ~75% which may explain the narrow size distribution observed in for $Fe_{0.75}Co_{0.25}$ Therefore, the catalytic performance of Fe-Co catalysts may be governed by a trade-off between maintaining high-temperature structural integrity and providing active sites with sufficient reactivity.

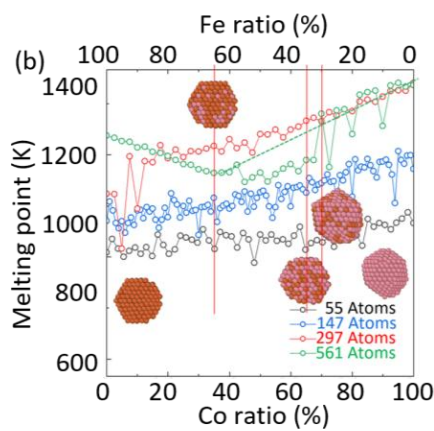

**Figure 6.** Melting point of different size Fe-Co nanoparticles obtained via MLFF driven MD simulations of slow cooling followed by slow heating at 10 K ns$^{-1}$.

**CONCLUSION**

This study systematically investigates the role of $Fe_xCo_{1-x}$ catalyst composition in SWCNT growth behavior via ACCVD. By combining optical spectroscopies, TEM analysis, and MD simulations, we demonstrate that both catalyst composition and growth conditions considerably influence SWCNT yield, diameter/chirality distribution, and quality. Key findings reveal a diameter-dependent growth trend governed by catalyst composition and growth temperature. At 600 °C, $Fe_0Co_1$ catalysts produce the highest SWCNT yield, with diameters predominantly concentrated in the range of 0.7–0.9 nm. In contrast, at 850 °C, Fe-rich catalysts ($Fe_{0.75}Co_{0.25}$) exhibit superior performance, favoring relatively larger-diameter SWCNTs (0.9–1.1 nm). TEM and EDS analyses attribute this trend to the formation uniformly sized Fe-Co alloy NPs with surface segregated Co. MD simulations further support these observations, showing similar structures and high melting points of large $Fe_xCo_{1-x}$. This work highlights the importance of tailored $Fe_xCo_{1-x}$ catalyst design for achieving diameter-selective, high-yield SWCNT synthesis, offering practical insights for scalable and controlled nanotube production.

## METHODS

### Synthesis of SWCNTs on zeolite

Catalytic Fe and Co were supported on USY-zeolite (HSZ-390HUA, Tosoh) via impregnation with a homogeneous mixture of iron(II) acetate [$(CH_3COO)_2Fe$], cobalt(II) acetate tetrahydrate [$(CH_3COO)_2Co \cdot 4H_2O$] and ethanol, following a procedure analogous to that reported in previous studies.[16] Zeolite coated with catalytic Fe and Co NPs at varying molar ratios was evenly distributed on a quartz boat, which was then placed inside a quartz tube (27 mm in diameter) within an electric furnace, as shown in **Fig. S1a**. $Fe_xCo_{1-x}$, the ratio is the molar ratio of Co atoms to Fe atoms in the mixture of $(CH_3COO)_2Co \cdot 4H_2O$ and $(CH_3COO)_2Fe$, the total molar amount of metal atoms is fixed. During the heating process to the target temperature and the subsequent reduction reaction, 300 sccm of $Ar/H_2$ gas (3% $H_2$) was continuously introduced into the quartz tube, maintaining an internal pressure of 40 kPa. Following the reduction process, the $Ar/H_2$ gas flow was stopped, and the quartz tube was evacuated using a rotary pump. 310 sccm mixture gas of Ar and ethanol vapor, with Ar acting as the carrier gas, was then introduced to keep the target partial pressure of ethanol (typically 50Pa). Upon completing the CVD reaction, the system was cooled to room temperature under a flow of either Ar or Ar/H$_2$ gas. The growth temperature ranged from 500 °C to 900 °C, and the growth pressure (ethanol partial pressure unless otherwise specified) varied between 10 Pa and 10 kPa. After cooling, the SWCNT sample grown on zeolite in the quartz boat was collected for subsequent characterizations. For detailed information on CVD conditions, please refer to **Fig. S1a** and the previous reports.[16, 33, 37-39]

**Synthesis of SWCNTs on TEM grid**

Fe and Co catalysts were directly deposited onto a SiO$_2$/Si TEM grid (20 nm SiO$_2$ on the top) using magnetron sputtering (Advance Riko). The grid was then annealed in air at 400 °C for 5 min to fix the catalyst NPs. After annealing, the grid was transferred into a CVD chamber for SWCNT synthesis, as shown in **Fig. S1c**. After CVD growth, the grid, containing the catalysts only or the catalysts with as-grown SWCNTs, was used directly for TEM characterizations, enabling observation of the original morphology of the catalysts and SWCNTs. TEM characterizations were performed using a JEM-ARM200F microscope for atomic-resolution images and a JEOL 2010F microscope for general TEM imaging, both operated at an acceleration voltage of 200 kV. High-angle annular dark-field scanning TEM (HAADF-STEM) images were captured using the JEM-ARM200F microscope equipped with a cold field-emission gun, employing a probe size typically smaller than 0.1 nm. For additional details on direct observations using a SiO$_2$/Si TEM grid, please refer to the previous reports.[27, 36, 37]

**CHARACTERIZATIONS**

Samples were analyzed using a range of techniques. The solid samples were examined via scanning electron microscopy (SEM, Hitachi S-4800) at an accelerating voltage of 1 kV and a Renishaw inVia Raman system, employing laser excitation wavelengths of 488, 532, 633, and 785 nm. Additionally, dispersed SWCNT solution samples were prepared for absorption and photoluminescence-excitation (PLE) measurements. To prepare these solution samples, 8 mg of as-grown SWCNTs on zeolite powder were dispersed in 5 mL of a DOC/D$_2$O solution (0.5 wt% sodium deoxycholate (DOC) in deuterium oxide (D$_2$O)) through tip-sonication (Power: 400 W; Amplitude: 50%; Time: 15min) and centrifugation (The relative centrifugal force (RCF) was 11,180 × $g$; and time is 3 hours), as illustrated in **Fig. S1b**. The UV-vis-NIR absorbance and PLE spectra of the SWCNT solutions were recorded using a UV-3150 spectrophotometer (Shimadzu Co., Ltd.) and a HORIBA Jobin Yvon Fluorolog iHR320 equipped with a liquid-nitrogen-cooled InGaAs detector, respectively. To minimize reabsorption effects, the absorbance of the suspensions used for PLE measurements was kept below 0.1 (using a 1cm pathlength cuvette).

## DATA PROCESSING

### Background subtraction of absorption spectra

The collected raw SWCNT samples inevitably contain some impurities, such as amorphous carbon and fullerenes. These impurities can interfere with the subsequent optical characterizations. Besides, scattering effects also contribute to a background in absorption spectrum given the SWCNTs have a length in the range of the wavelength of the excitation light. As a result, the measured absorbance includes a strong background, indicated by the grey area in **Fig. 1b**. Therefore, prior to comparing the absorption spectra of different samples, it is essential to perform a proper background correction to eliminate the influence from the scatterings and the impurities. In this work, we adopted a background subtraction strategy reported by Tian *et al,*[40, 41] employing a combined Fano and Lorentzian function model to accurately correct the background in the absorption spectra, thereby ensuring the reliability and comparability of the data. The post-processed data are shown in **Fig. 1c** and **1d**, and a detailed description of the fitting process can be found in SI **Section 2**.

**Determination of chirality distribution**

Quantifying the relative abundance of each chiral species in SWCNTs synthesized with different $Fe_xCo_{1-x}$ catalysts is crucial to elucidate the influence of bimetallic composition on controlled growth. Two-dimensional photoluminescence excitation spectroscope (2D PLE) is a powerful technique for characterizing semiconducting SWCNTs. Since the optical emission and excitation transitions of SWCNTs are strongly correlated with their chirality, the peak positions in PLE spectra enable precise identification of specific chiral species. However, PLE spectra alone cannot directly provide the relative abundance of each chirality species due to variations in emission efficiencies, which depend on both chirality and diameter,[42] as well as environmental interactions.[43] Still, comparing PLE spectra measured in the same manner can provide relative changes in chiral species abundancy between samples. Likewise, absorption spectra cannot provide absolute quantification of chiral species abundance due to diameter- and environment-dependent absorption cross-sections, however they remain a direct indicator of relative abundance for each (n, m) species. Moreover, because overlapping peaks in the absorption spectra complicate the analysis, the data can only provide an approximate estimate of the overall chirality distribution. Since neither PLE nor absorption spectra alone can accurately estimate chiral species abundance, we combined both methods to achieve a more reliable quantification. Firstly, we systematically fitted the 2D PLE spectra using the method developed by Cambré and co-workers,[43] extracting key spectral parameters such as peak positions and full width at half maximum (FWHM) for each chirality. These parameters were then used to deconvolute the absorption spectra, enabling reliable quantification of the relative abundance of each chiral species (**Fig. 2a and 2c**). Further methodological details are available in the SI **Section 2**.

**MOLECULAR DYNAMICS SIMULATION**

Molecular dynamics (MD) simulations were carried out using the Graphics Processing Units Molecular Dynamics (GPUMD) software[1], version 3.9.5. To determine the structure and melting point of each cluster size, simulations were performed in two stages.

In the first stage, each cluster was cooled from its initial temperature to 300 K at a rate of 10 K ns$^{-1}$. The initial temperatures were chosen based on cluster size: 1300 K for $M_{55}$, 1400 K for $M_{147}$, 1500 K for $M_{297}$, and 1600 K for $M_{561}$. After cooling, each cluster was relaxed to obtain its equilibrium structure. In the second stage, the relaxed clusters were heated from 300 K back to their respective initial temperatures—1300 K for $M_{55}$, 1400 K for $M_{147}$, 1500 K for $M_{297}$, and 1600 K for $M_{561}$—again at a rate of 10 K ns$^{-1}$. The Lindemann index[2] was calculated during the heating simulations to determine the melting point. A cluster was considered to have melted when the Lindemann index exceeded 0.21.

All MD simulations used a timestep of 2.0 fs and were conducted in the NVT ensemble, with temperature controlled by a Langevin thermostat[3] using a coupling constant of 100. At each timestep, both linear and angular momentum were removed to prevent drift and rotation. A Fe-Co neuroevolution potential[4-7] (NEP) under development was employed to model interatomic interactions. While the details of this potential will be published separately, its reliability was assessed during the simulations by using an ensemble of NEPs to estimate the model deviation (i.e., the variation in predicted forces). For each melting simulation, the mean model deviation was below 250 meV/Å, which is considered acceptable for the current study. All structures obtained from the MD simulations was visualized using the OVITO software[8].

**AUTHOR INFORMATION**

Corresponding Author

**Shigeo Maruyama** − Department of Mechanical Engineering, The University of Tokyo, Tokyo 113-8656, Japan; State Key Laboratory of Fluid Power and Mechatronic Systems, School of Mechanical Engineering, Zhejiang University, Hangzhou 310027, China; orcid.org/0000-0003-3694-3070; Email: maruyama@photon.t.u-tokyo.ac.jp

**Ya Feng**− Key Laboratory of Ocean Energy Utilization and Energy Conservation of Ministry of Education, School of Energy and Power Engineering, Dalian University of Technology, Dalian, Liaoning 116024, China, Email: fengya@dlut.edu.cn

Authors

**Qingmei Hu**−Department of Mechanical Engineering, The University of Tokyo, Tokyo 113-8656, Japan

**Wanyu Dai**− Department of Mechanical Engineering, The University of Tokyo, Tokyo 113-8656, Japan

**Daisuke Asa**−Department of Mechanical Engineering, The University of Tokyo, Tokyo 113-8656, Japan

**Daniel Hedman**−Center for Multidimensional Carbon Materials (CMCM), Institute for Basic Science (IBS) (Republic of Korea), Ulsan, 44919, Republic of Korea

**Aina Fitó-Parera**−Theory and Spectroscopy of Molecules and Materials, Department of Physics, University of Antwerp, Antwerp 2610, Belgium

**Yixi Yao**−Center Beijing National Laboratory for Molecular Science, Key Laboratory for the Physics and Chemistry of Nanodevices, State Key Laboratory of Rare Earth Materials Chemistry and Applications, College of Chemistry and Molecular Engineering, Peking University, Beijing, 100871, China

**Yongjia Zheng**−State Key Laboratory of Fluid Power and Mechatronic Systems, School of Mechanical Engineering, Zhejiang University, Hangzhou 310027, People's Republic of China; orcid.org/0000-0001-5836-6978

**Kaoru Hisama**−Interdisciplinary Cluster for Cutting Edge Research, Research Initiative for Supra-Materials, Shinshu University, Japan

**Christophe Bichara**−The Interdisciplinary Nanoscience Centre of Marseille (CINaM), Aix-Marseille University and CNRS, Marseille, France

**Shohei Chiashi**−Department of Mechanical Engineering, The University of Tokyo, Tokyo 113-8656, Japan; orcid.org/0000-0002-3813-0041

**Yan Li**−Center Beijing National Laboratory for Molecular Science, Key Laboratory for the Physics and Chemistry of Nanodevices, State Key Laboratory of Rare Earth Materials Chemistry and Applications, College of Chemistry and Molecular Engineering, Peking University, Beijing, 100871, China

**Wim Wenseleers**−Nanostructured and Organic Optical and Electronic Materials, Department of Physics, University of Antwerp, Antwerp 2610, Belgium

**Dmitry I. Levshov**−Theory and Spectroscopy of Molecules and Materials, Department of Chemistry, University of Antwerp, Antwerp 2610, Belgium; orcid.org/0000-0002-2249-7172

**Sofie Cambré**−Theory and Spectroscopy of Molecules and Materials, Department of Physics, University of Antwerp, Antwerp 2610, Belgium; orcid.org/0000-0001-7471-7678

**Keigo Otsuka**−Department of Mechanical Engineering, The University of Tokyo, Tokyo 113-8656, Japan; orcid.org/0000-0002-6694-0738

**Rong Xiang**− Department of Mechanical Engineering, The University of Tokyo, Tokyo 113-8656, Japan; State Key Laboratory of Fluid Power and Mechatronic Systems, School of Mechanical Engineering, Zhejiang University, Hangzhou 310027, People's Republic of China; orcid.org/0000-0002-4775-4948

**Funding Sources**

This work was financially supported by the Japan Society for the Promotion of Science (JSPS) KAKENHI under Grant Numbers JP24KJ0751, JP23H00174, JP23H05443, and JP21KK0087, as well as by JST CREST under Grant Number JPMJCR20B5, Japan. Additionally, this work was supported by a JSPS-FWO Bilateral Joint Research Project (Grant Numbers JSPS JPJSBP120212301 and FWO VS08521N). Y.F. acknowledges support from the National Natural Science Foundation of China (No. 524766164), The Talent Development Program for Liaoning Province (No. XLYC2403036), as well as Science and Technology Innovation Fund of Dalian, International Science and Technology Cooperation (No. 2024JJ12RC035). A.F.P. acknowledges the Research Foundation Flanders (FWO) through a personal PhD fellowship (1178324N). R.X. acknowledges the support by the National Key R&D Program of China (2024YFA1409600, 2023YFE0101300) from the Ministry of Science and Technology of China and research fund from Zhejiang province (2022R01001).

**Notes**

The authors declare no competing financial interest.

**ABBREVIATIONS**

SWCNT: Single-Walled Carbon Nanotube

Iron: Fe

Cobalt: Co

CVD: Chemical Vapor Deposition

MD: Molecular dynamics

CNT: Carbon Nanotube

NPs: Nanoparticles

MWCNT: Multi-Walled Carbon Nanotube

TEM: Transmission electron microscopy

HAADF-STEM: High-angle annular dark-field scanning TEM

SEM: Scanning electron microscopy

PLE: Photoluminescence-excitation

DOC: Sodium Deoxycholate

$D_2O$: Deuterium oxide

2D PLE: Two-dimensional photoluminescence-excitation

FWHM: Full width at half maximum

ACCVD: Alcohol Catalytic Chemical Vapor Deposition

TGA: Thermogravimetric analysis

# Supplementary Information

**1. Supplementary figures for the discussion in the main text**

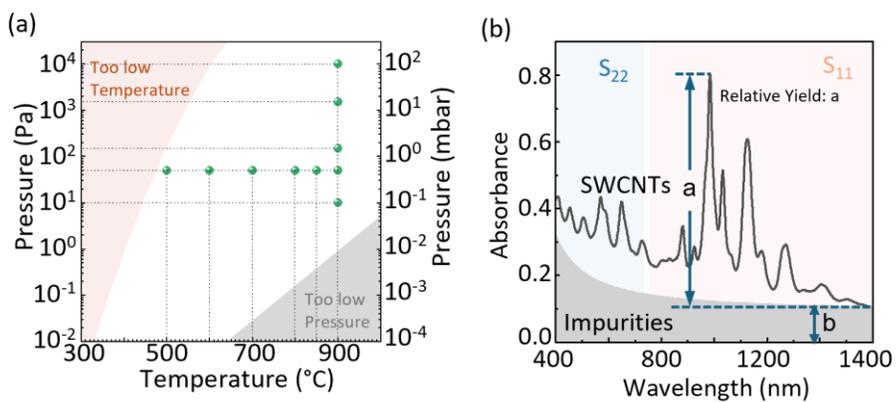

**Figure S1.** Study of the growth efficiency of SWCNTs through ACCVD. (a) The broad operational window of ACCVD.[1] (b) A typical baseline subtraction procedure of UV-vis-NIR absorption spectrum using the function model of Fano and Lorentzian developed by Tian *et al.*[2]

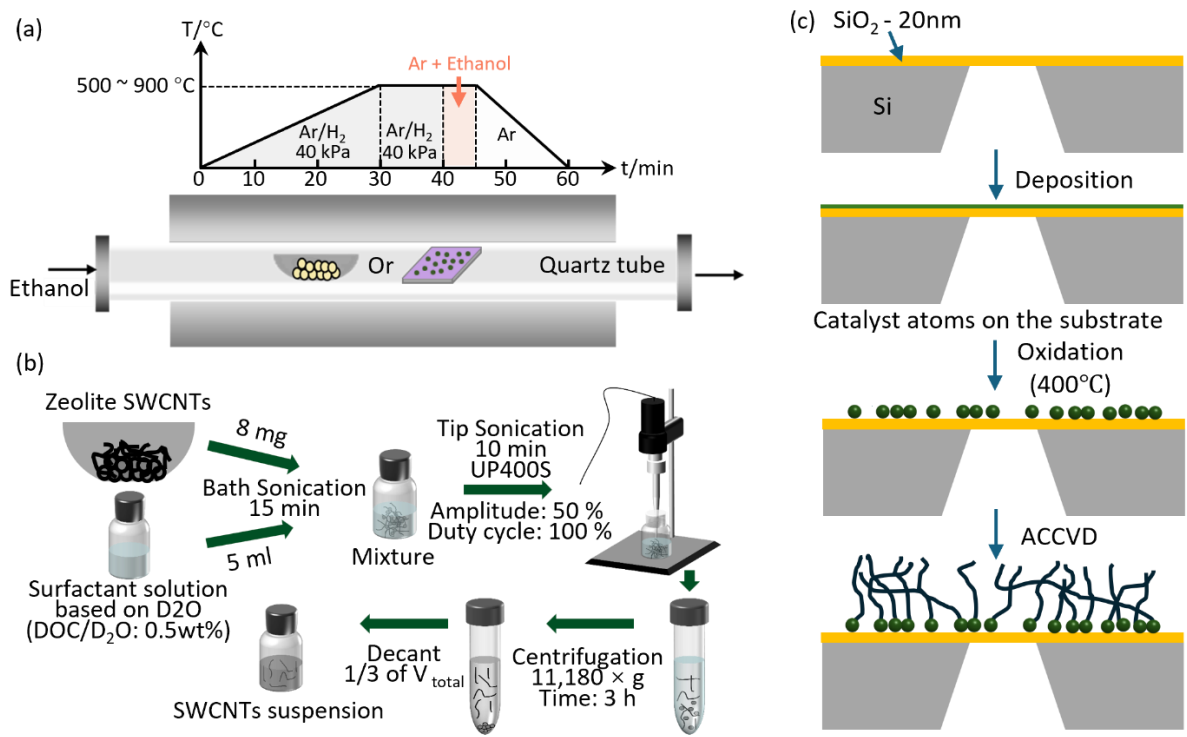

**Figure S2.** The sample preparation procedure. ACCVD for (a) zeolite-supported or (c) $SiO_2$/Si-supported SWCNT growth; (b) Dispersion processes for preparing suspension of SWCNT.

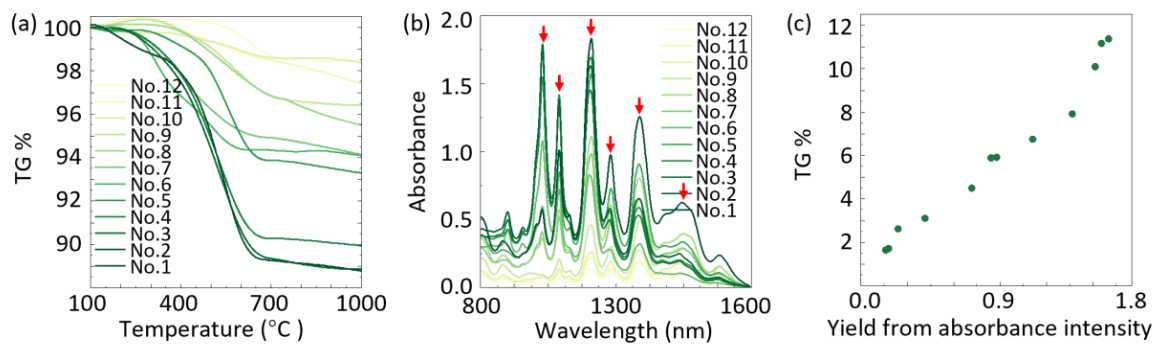

**Figure S3.** Verify the relationship between TGA and Absorbance. (a) TGA analysis and (b) Absorption spectra (using a 1cm pathlength cuvette.) of the samples obtained by 12 various sets of growth parameters; The absorption peaks indicated by the red arrows are selected, and their intensities are summed to generate the data presented in Figure c. (c) The dependence of TGA weight loss with respect to absorbance, showing a very nice liner dependence. Notably, zeolite represents the dominant component in the original sample, with SWCNTs comprising only around 10% of the total weight.

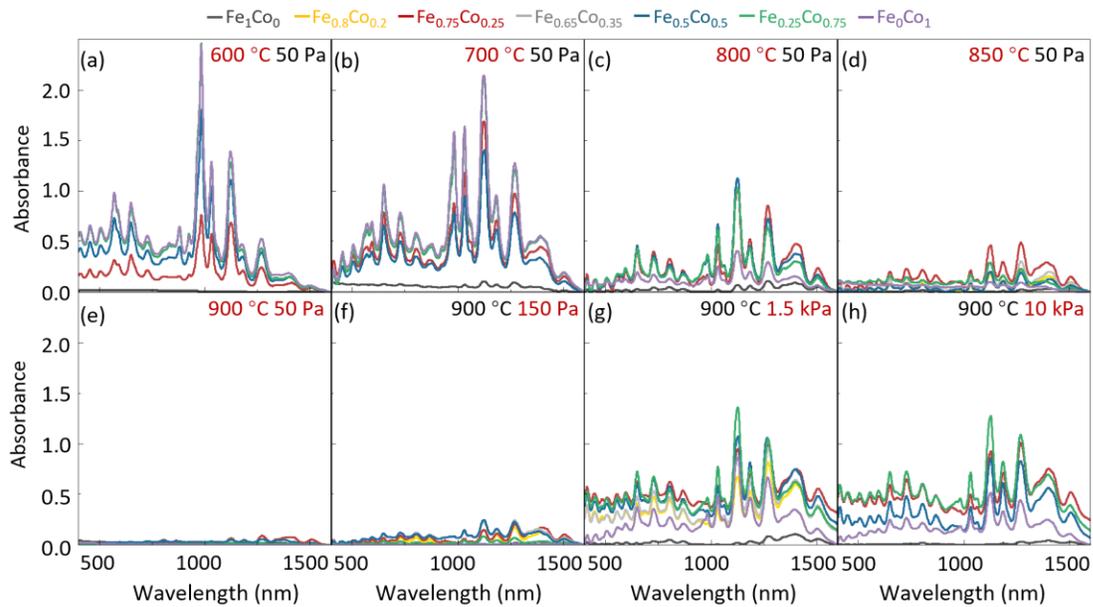

**Figure S4.** The absorption spectra for samples prepared by different growth conditions, with varying Fe-Co ratios at different temperatures or pressure.

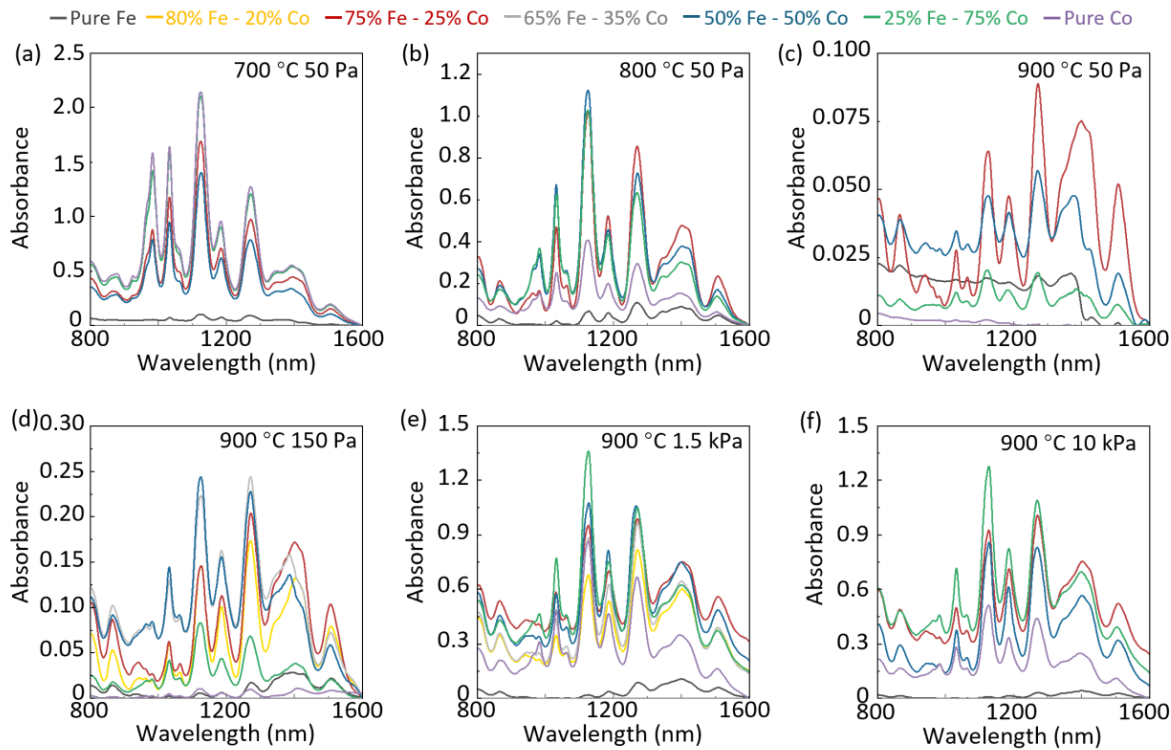

**Figure S5.** The absorption spectrum of samples grown at (a) 700 °C 50 Pa; (b) 800 °C 50 Pa; (c) 900 °C 50 Pa; (d) 900 °C 150 Pa; (e) 900 °C 1.5 kPa. Note the different y-scales of the panels.

**Fig. S4** and **S5** present the absorbance spectra of two sets of SWCNT samples synthesized under varying growth conditions, including different Fe/Co atomic ratios in the catalysts, growth temperatures, and ethanol partial pressures. To enable a direct visual comparison across all conditions, each subplot in **Fig. S4** employs a unified coordinate range. As shown in **Fig. S4**, under 600 °C growth conditions, all catalysts except pure Fe exhibit high absorbance values, indicating generally high growth efficiency. In contrast, when the temperature increases to 900 °C while maintaining a low ethanol pressure of 50 Pa, absorbance values significantly decrease, with some samples falling below 0.1 (**Fig. S4e**). This suggests that such conditions approach the practical limit of growth, making reliable optical analysis infeasible. Conversely, at 900 °C, increasing the ethanol partial pressure leads to a marked enhancement in absorbance (**Fig. S4e**–**S4h**), highlighting the critical role of pressure in high-temperature growth. This trend aligns with previous studies showing that in ACCVD systems, both relatively low-temperature/low-pressure and relatively high-temperature/high-pressure conditions are favorable for SWCNT synthesis, while relatively high-temperature/low-pressure conditions represent a more stringent and less efficient growth regime. To improve the clarity of individual spectra, the vertical axis range in each subplot was independently optimized in **Fig. S5**, allowing better visualization of variations in absorbance among samples.

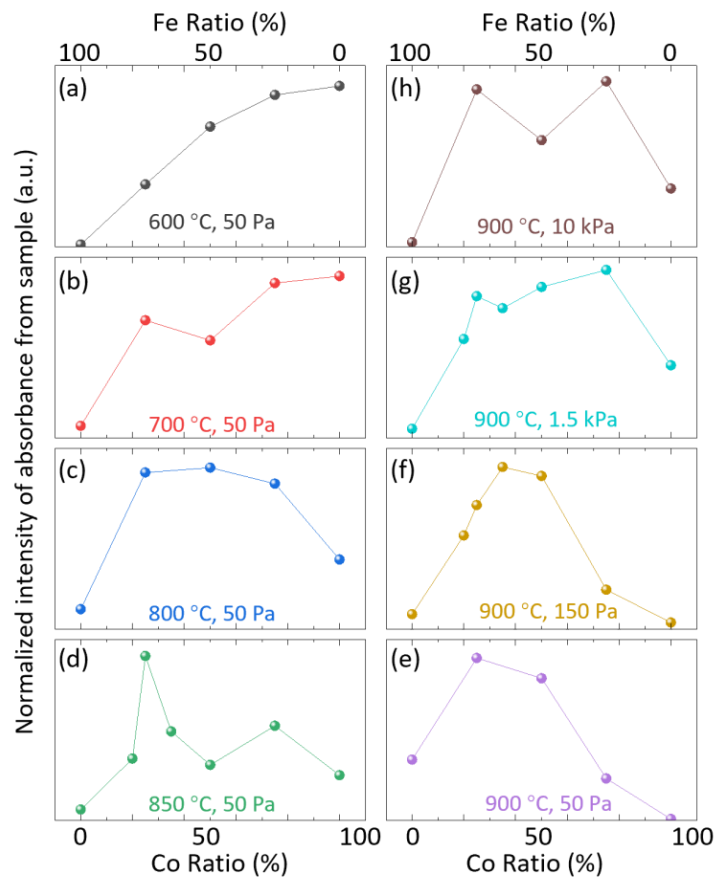

**Figure S6.** Effect of atomic ratio in bimetallic catalysts on the absorbance of SWCNTs grown under different ACCVD conditions: (a-e) different growth temperatures; and (e-h) different growth pressure.

As previously discussed, absorbance serves as a reliable indicator of SWCNT growth efficiency. To systematically evaluate the catalytic performance of Fe/Co bimetallic catalysts under various conditions, the absorbance spectra of individual samples were further processed (similar to **Fig. S3b**) to extract representative absorbance values. These average values were then used to construct **Fig. S6**, which reveals the correlation between average absorbance and catalyst composition.

In the first series of experiments (**Fig. S6a–S6e**), conducted at a constant ethanol pressure of 50 Pa, the highest absorbance was observed for samples synthesized with pure Co catalysts at 600 °C, while those using pure Fe showed negligible activity. As the temperature increased, the optimal Fe/Co ratio gradually shifted toward higher Fe content, indicating that Fe atoms began to participate in the catalytic process at elevated temperatures, thereby exhibiting a synergistic effect between Fe and Co. At 850 °C, the $Fe_{0.75}Co_{0.25}$ composition already showed the highest absorbance, which was further confirmed as the optimal catalyst at 900 °C, supporting the notion that increased Fe content enhances growth efficiency under these conditions.

In contrast, when the growth temperature was fixed at 900 °C and the ethanol partial pressure was increased (**Fig. S6e–S6h**), the optimal composition shifted toward higher Co content. At a pressure of 1.5 kPa, $Fe_{0.25}Co_{0.75}$ emerged as the most effective catalyst. Based on the results shown in **Fig. S3** and **S6**, it can be inferred that under harsher growth conditions (relatively, high temperature, low pressure), Fe plays a more crucial role in sustaining growth, whereas under more favorable conditions (high temperature and high pressure or low temperature and low pressure, relatively), the catalytic activity is dominated by Co.

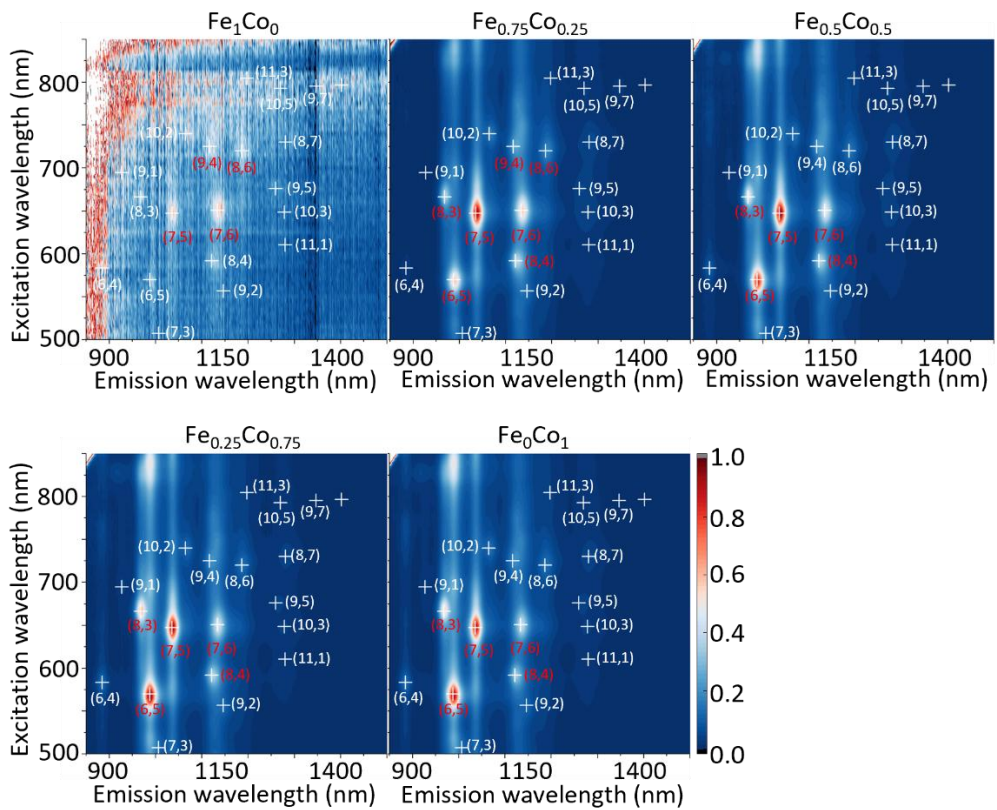

**Figure S7.** The normalized photoluminescence excitation maps of SWCNT dispersed solution samples grown at 600 °C.

**Figure S8.** The normalized photoluminescence excitation maps of SWCNT dispersed solution samples grown at 850 °C.

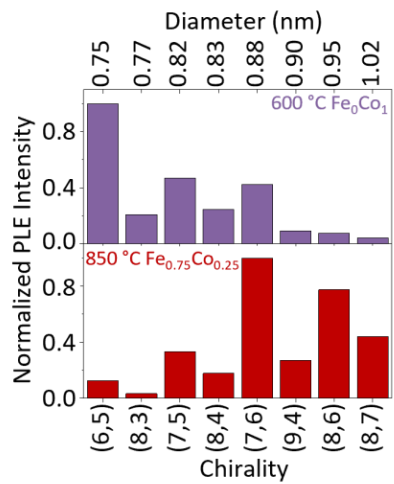

**Figure S9.** Normalized photoluminescence excitation intensity (from **Fig. S7** and **S8**) bar chart for SWCNTs grown using $Fe_0Co_1$ catalysts at 600 °C and $Fe_{0.75}Co_{0.25}$ catalysts at 850 °C. A clear shift in intensity is observed between the two growth conditions.

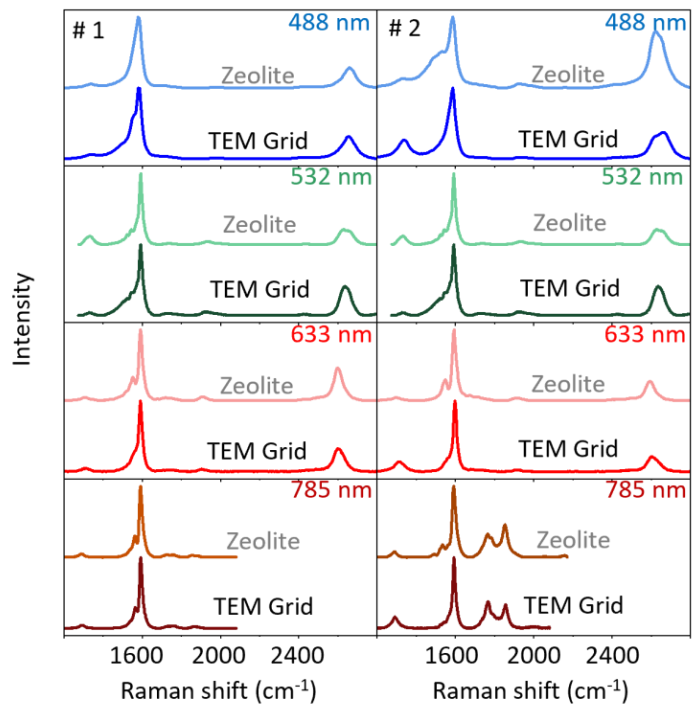

**Figure S10.** The Raman spectra (in the range of D, G and 2D bands) for two sets of samples grown on zeolite and SiO$_2$/Si TEM grid. Here #1 is Fe$_{0.5}$Co$_{0.5}$ and #2 is Fe$_0$Co$_1$, synthesized at 800 °C, 1.2kPa and 600 °C, 50Pa, respectively.

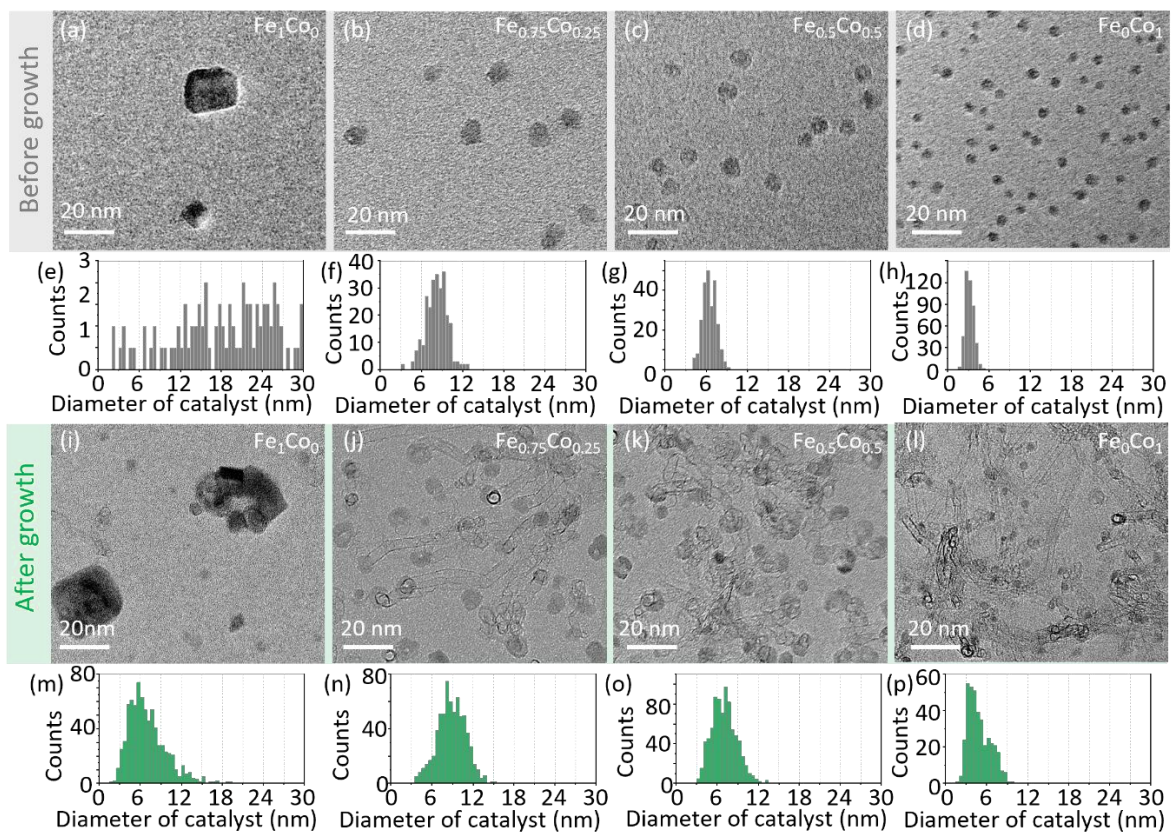

**Figure S11.** TEM characterization and analysis of samples grown at 600 °C and 50 Pa. TEM images of samples with different ratio before (a-d) and after (i-l) growth; Histogram of catalyst diameter for $Fe_xCo_{1-x}$ samples before (a-d) and after (i-l) growth.

In contrast to the data presented in **Figure 4** in the main text at 850 °C, here we present the data at 600 °C. Before the ACCVD process, the catalysts of $Fe_1Co_0$ exhibited a broad particle size distribution, with the majority of particles ranging between 20 and 30 nm and only a small fraction measuring below 10 nm (**Fig. S11a**). As the Co ratio increased, the particle size distribution shifted toward smaller diameters, with $Fe_0Co_1$ catalysts showing the narrowest distribution, centered around 3 nm (**Fig. S11d**). Following the ACCVD process (**Fig. S11i–S11p**), all samples except those with $Fe_1Co_0$ catalysts successfully induced SWCNT growth. Among the effective catalysts ($Fe_{0.75}Co_{0.25}$, $Fe_{0.5}Co_{0.5}$, and $Fe_0Co_1$), the density of SWCNTs increased with increasing Co ratio, with $Fe_0Co_1$ catalysts demonstrating the highest SWCNT density. Notably, the post-growth catalyst particle sizes aligned with the pre-growth trends, reinforcing the critical influence of catalyst composition, namely $Fe_xCo_{1-x}$ ratio, and consequently particle size, on catalytic performance under these ACCVD conditions.

## 2. Detailed fitting process and results of absorption spectra

The entire fitting procedure consists of two main parts: two-dimensional (2D) fitting of fluorescence-excitation (PLE) maps and absorbance (Abs) spectral fitting. The 2D fitting of PLE was performed using the fitting software developed by Cambré et al. From this analysis, we systematically extracted the peak positions (Emission wavelength (S11) and excitation wavelength (S22)), full width at half maximum (FWHM), and other relevant parameters for each chirality, which were subsequently converted to the parameters for subsequent absorbance fitting (the detailed conversion procedure is described in **Section 2.3**). The converted parameters were then applied in the peak fitting of the absorbance spectra (That is the pristine Abs. spectrum mentioned in **Fig. 4a** and **4c** in the main text), where Lorentzian functions were used as the fitting model. The detailed processes are described as follows.

### 2.1 PLE fitting

The experimentally obtained 2D PLE spectra were imported into the PLE fitting software. This tool uses a precise empirical model for the excitation line shape and integrate it with an emission line shape model within a 2D fitting framework, based on the model developed in reference.[3, 4] This approach enables accurate fitting of 2D PLE maps for various SWCNT samples, facilitating the direct extraction of line shape features, including peak positions, linewidths, intensities, and other physical quantities, such as phonon sidebands in both excitation and emission spectra. During the fitting process, appropriate parameter settings and meticulous validation ensured the accuracy and reliability of the fitting results.

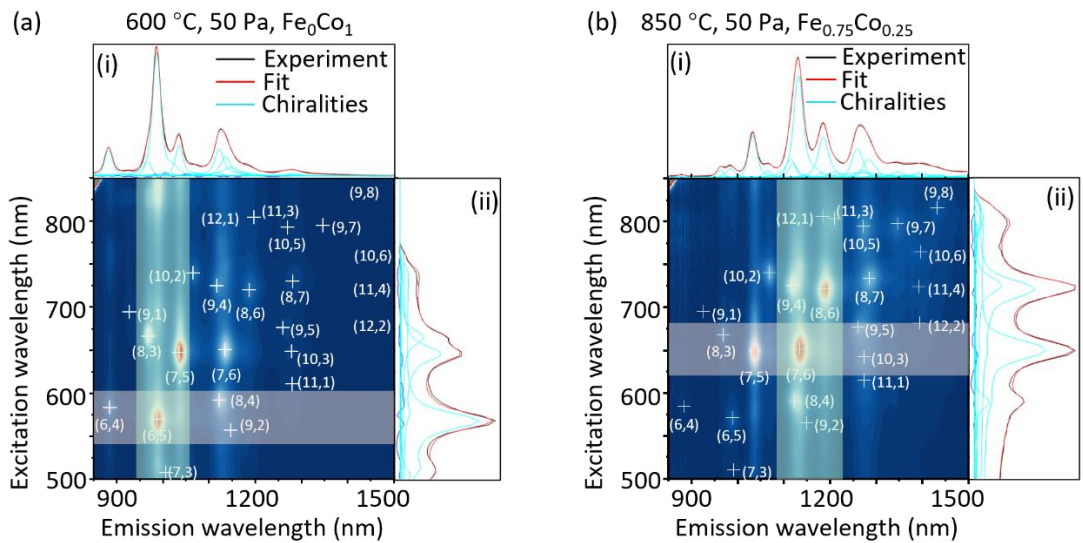

**Figure S12.** The PLE fitting window of (a) $Fe_0Co_1$ grown at 600 °C 50 Pa and (b) $Fe_{0.75}Co_{0.25}$ grown at 850 °C 50 Pa. Among, (i) fitted emission spectra and (ii) fitted excitation spectra showing the contributions from all chiral species (in cyan).

As shown in **Fig. S12**, the fitting panel provides an intuitive interface that enables users to select specific spectral regions of interest by using a rectangular selection cursor. This functionality facilitates focused and localized evaluation of the fitting results, allowing users to closely inspect individual spectral segments where complex peak overlaps or baseline variations may occur. Furthermore, as illustrated in **Fig. S13**, the overall fitting quality can be comprehensively assessed by comparing the spot size, shape, and precise position of chiral peaks in the fitted spectrum with those in the original experimental data. This comparative analysis ensures that both the intensity distribution and peak resolution are well reproduced by the fitting algorithm.

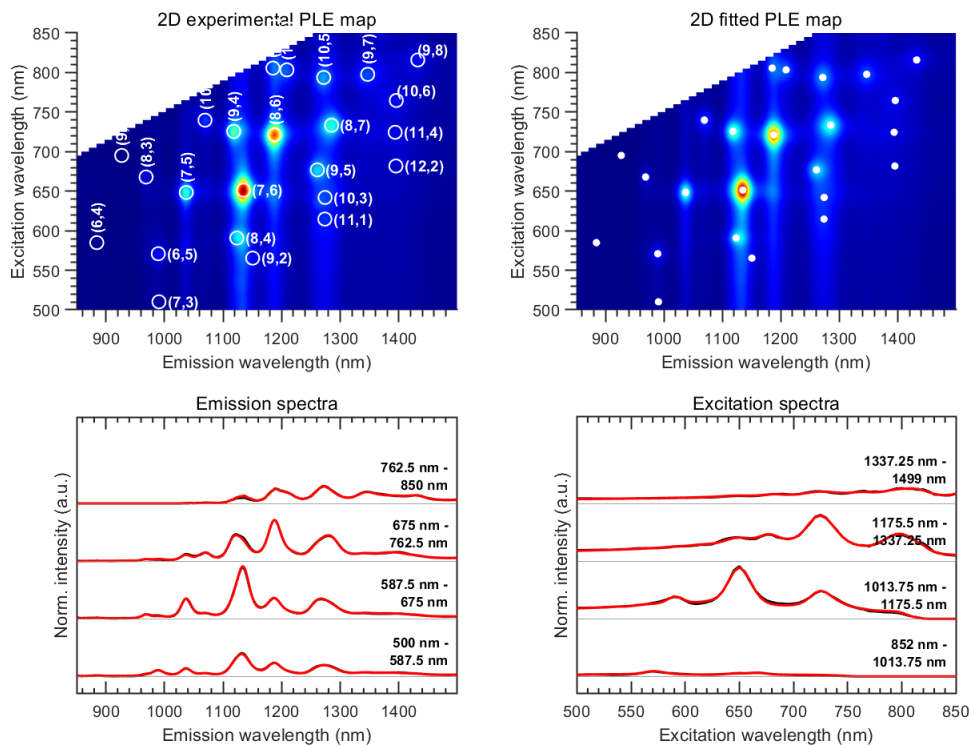

**Figure S13.** Experimental and fitted PLE maps showing excitation and emission spectra by integrating over specific regions of the larger PLE maps, where experimental data (black) and fitted data (red) nicely overlap. The example shown here is for the $Fe_{0.75}Co_{0.25}$ sample grown at 850 °C 50 Pa. Fitted peaks are indicated by the white markers in the top right panel, while corresponding chiral indices are given in the left top panel.

**2.2 Absorption spectrum fitting**

The peak positions, linewidths, intensities, and phonon sideband parameters obtained from the PLE fitting were converted accordingly and used as initial input parameters for the peak fitting of the absorbance spectra. Lorentzian functions were applied as the fitting model. The final fitting results are presented in **Fig. 2a** and **2c**.

**2.3 Derivation of the relationship between PLE fitting parameters and those required for absorption fitting**

The core of this fitting strategy lies in the rational and effective utilization of the parameters obtained from PLE fitting. To achieve this, a series of systematic verifications were conducted. Initially, the peak positions and linewidths obtained from PLE fitting were directly applied as fixed parameters in the absorbance fitting process. As shown in **Fig. S14a**, comparison of the fitted and original spectra at the (7,5) and (6,5) peaks revealed a noticeable shift of approximately 4 nm towards the blue. This observation indicates that directly using PLE-derived peak positions in the absorbance fitting process leads to systematic deviations, however they can serve as ideal starting parameters for optimizing the absorption fits after applying a general shift, as the relative intensities in the absorption spectrum can already be quite nicely reproduced. Further analysis across multiple samples confirmed the consistency of this shift between absorption and emission peaks, with the degree of shift closely correlated with the type of dispersant used for SWCNT dispersion. The statistical results for different dispersants and organic polymers, including Sodium deoxycholate (DOC), Poly(9,9-dioctylfluorenyl-2,7-diyl)-alt-2,2′-bipyridine (PFO-BPy), Poly(N-vinylcarbazole) (PCz), and P1 (developed by Prof. Li Yan's group at Peking University), are summarized in Table 1. These findings highlight the necessity of shifting the peak positions obtained from PLE fitting before applying them in absorbance fitting. Interestingly, even when applying a single general fixed shift of all the peaks a noticeable improvement in fitting accuracy can be observed as shown in **Fig. S14b**, though such a fixed shift does not fit all the data extremely accurately.

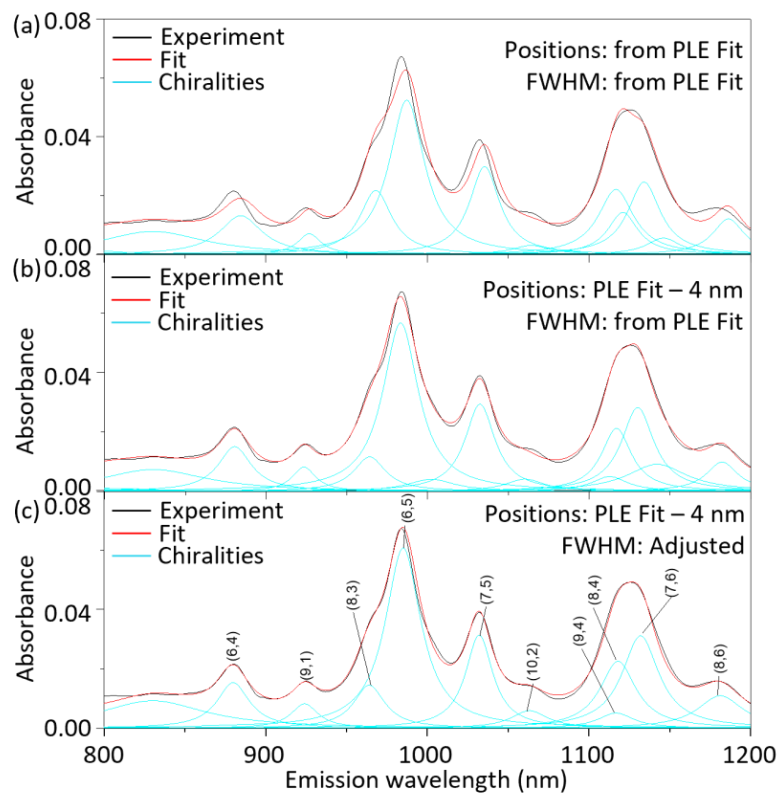

**Figure S14.** Comparison of different fit parameters through the same absorbance spectrum between 800 -1200nm.

Subsequently, we explored the conversion relationship between the linewidth parameters derived from PLE fitting and those required for absorbance fitting. Using the peak positions as fixed parameters, we manually adjusted the linewidths with reference to the PLE fitting results to optimize the absorbance fitting. Note that in the PLE fitting program, a Voigt line shape with corresponding FWHM was used, while here we use Lorentzian line shapes, thus therefore the required adaptation of the line width is logical. The final fitting outcome is presented in **Fig. S14c**, showing a very nice correspondence between experiment and fit, without requiring to adjusting all peak fitting parameters for each of the individual chiralities which would result in too many fit parameters to realistically optimize. Additionally, the peak strength ratio of S11 to S22 of each chiral peak obtained by fitting was calculated and plotted as **Fig. S15.** The peak intensity ratio of S11 to S22 for most chirality is found to be clustered around 3.1. This further supports the reliability of our fitting strategy, given it is expected that the ratio should be constant.

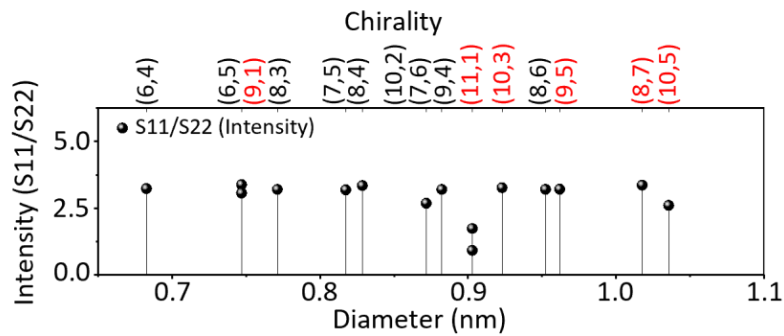

**Figure S15.** Intensity ratio of S11 and S22 for different chiralities as a function of SWCNT diameter.

The extracted linewidth values at this stage were compared with those obtained from PLE fitting, revealing a clear linear correlation between the two sets of parameters, as illustrated in **Fig. S16**. Notably, the differences in emission peak positions between the PLE fitting and absorption fitting results remain consistent, with a stable offset of approximately 4 nm. This indicates that applying an appropriate and fixed wavelength correction is essential prior to absorption fitting in the S11 region. In contrast, the differences in excitation peak positions between PLE fitting and absorption fitting are essentially negligible, suggesting that no wavelength shift occurs between the two fitting methods in the S22 region.

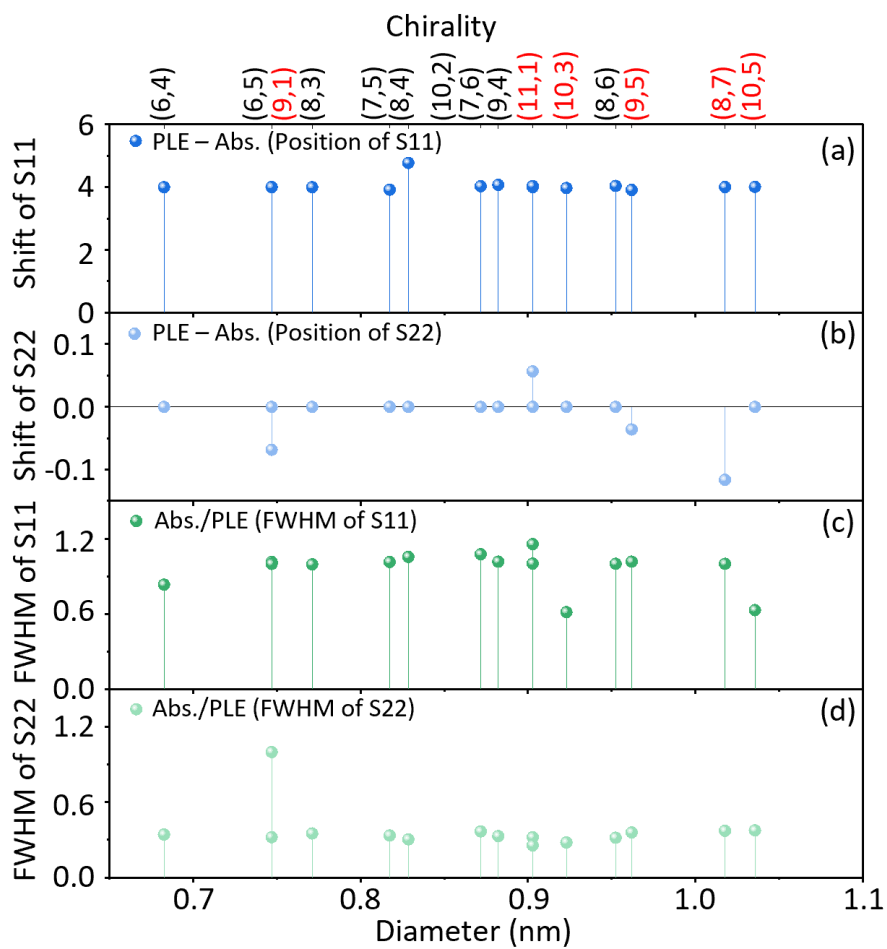

**Figure S16.** The extracted parameters of S11 and S22 for different chirality. Shift of the positions of (a) S11 and (b) S22; The ratio of FWHM of (c) S11 and (d) S22.

To further assess whether the fitting parameters are influenced by the choice of dispersants used for SWCNT dispersion, we dispersed the same SWCNT sample using different dispersants and performed identical optical characterizations, including absorption spectroscopy and PLE mapping. The fitting procedures were conducted following the same methodology. As shown in **Fig. S17**, the fitting results for the sample dispersed with Poly[(9,9-dioctylfluorenyl-2,7-diyl)-alt-(2,2'-bipyridine-5,5')] (PFO-BPy) are presented. Like the DOC-dispersed system, a consistent emission peak offset between PLE and absorption fitting results was observed. However, the offset for the PFO-BPy system was approximately −8 nm, differing in magnitude and sign from the DOC case. In contrast, the S22 excitation peak positions showed negligible differences between PLE and absorption fitting results, again consistent with the trend observed for DOC. These observations collectively demonstrate that the emission peak offset between the two fitting approaches is primarily induced by the dispersant environment, with the offset value depending on the dispersant type. A systematic summary of the emission peak offsets for SWCNTs dispersed in various dispersants is provided in **Table 1** (Readers may refer to this table when performing absorption fitting in subsequent analyses.).

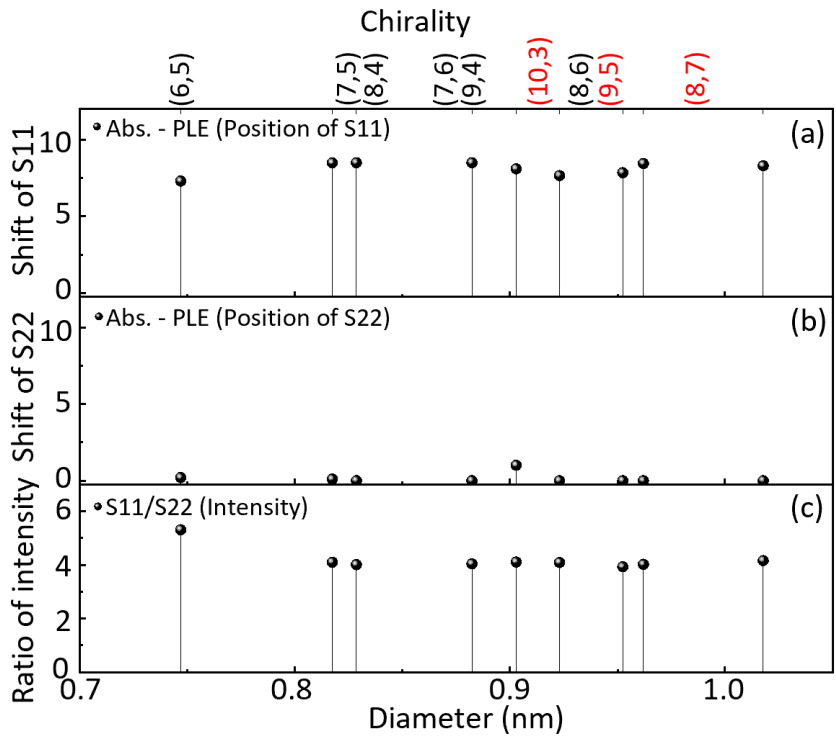

**Figure S17.** The extracted parameters of S11 and S22 for different chirality based on PFO-BPy as the dispersant. Shift of the positions of (a) S11 and (b) S22; (c) The ratio of intensity of S11 and S22.

**Table S1.** The summary of the recommended offset of S11 and S22 for the reference of Abs. fitting.

|   | DOC | PFO-BPy | PCz | P1@PKU |
|---|---|---|---|---|
| Shift of S11 (nm) (PLE – Abs.) | 4 | -8 | -7.4 | 7.5 |
| Shift of S22 (nm) (PLE – Abs.) | 0 | 0 | -0.6 | 0 |
| Ratio of FWHM for S11 (PLE / Abs.) | 0.95 | 1.29 | 1.23 | 1.15 |
| Ratio of FWHM for S22 (PLE / Abs.) | 0.31 | 0.57 | 0.61 | 0.5 |

## 2.4. Detailed fitting results of absorbance spectra for $Fe_0Co_1$ sample grown at 600 °C 50 Pa and $Fe_{0.75}Co_{0.25}$ sample grown at 850 °C 50 Pa

After establishing the fitting protocol and finalizing all fitting procedures, we proceeded to perform peak deconvolution of the absorption spectrum for the target samples grown under two distinct conditions: 600 °C at 50 Pa and 850 °C at 50 Pa. Based on the fitting results, empirical PL efficiency of different chiral species was calculated using the following equation:

$$\textit{Empirical PL Efficiency} = \frac{PL E\ Intensity}{Absorption\ Intensity}$$

For this, we use the (7,6) chirality as a normalized reference. Here, we focus on presenting and analyzing the results obtained for the two representative samples discussed in the main text, as shown in **Fig. S18** and **S19**. In the following sections, these two samples are subjected to further comparative analysis and detailed discussion, with particular attention given to the influence of growth temperature on chiral selectivity, peak resolution, and the variation in quantum efficiency across different nanotube species.

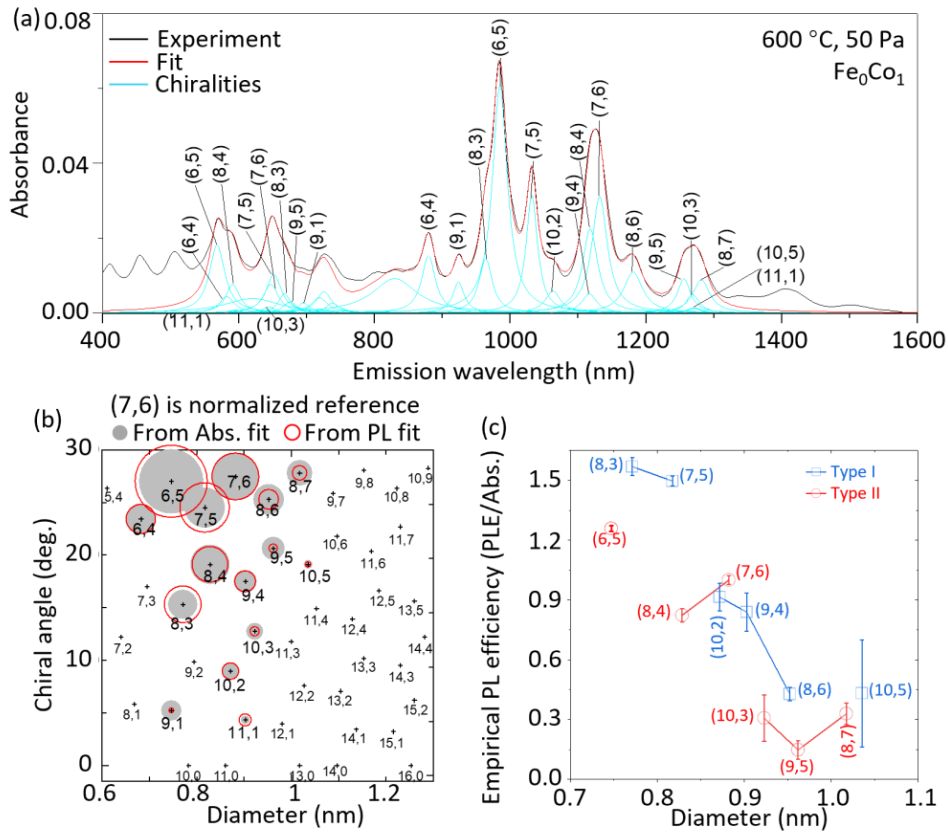

**Figure S18.** (a) Absorption spectrum and corresponding peak deconvolution results for SWCNTs grown at 600 °C and 50 Pa ($Fe_0Co_1$ catalyst). The experimental spectrum (black), overall fit (red), and individual chiral contributions (cyan) are shown. (b) 2D chirality distribution map plotted with tube diameter on the x-axis and chiral angle on the y-axis. The circle sizes represent relative intensities obtained from absorption fitting (gray) and PLE fitting (red outlines), with (7,6) used as a normalization absorption vs. PLE results. (c) Empirical PL efficiency (PLE/Abs) as a function of diameter for both Type I and Type II chiralities, highlighting the higher PL efficiency of small-diameter species. [5-7]

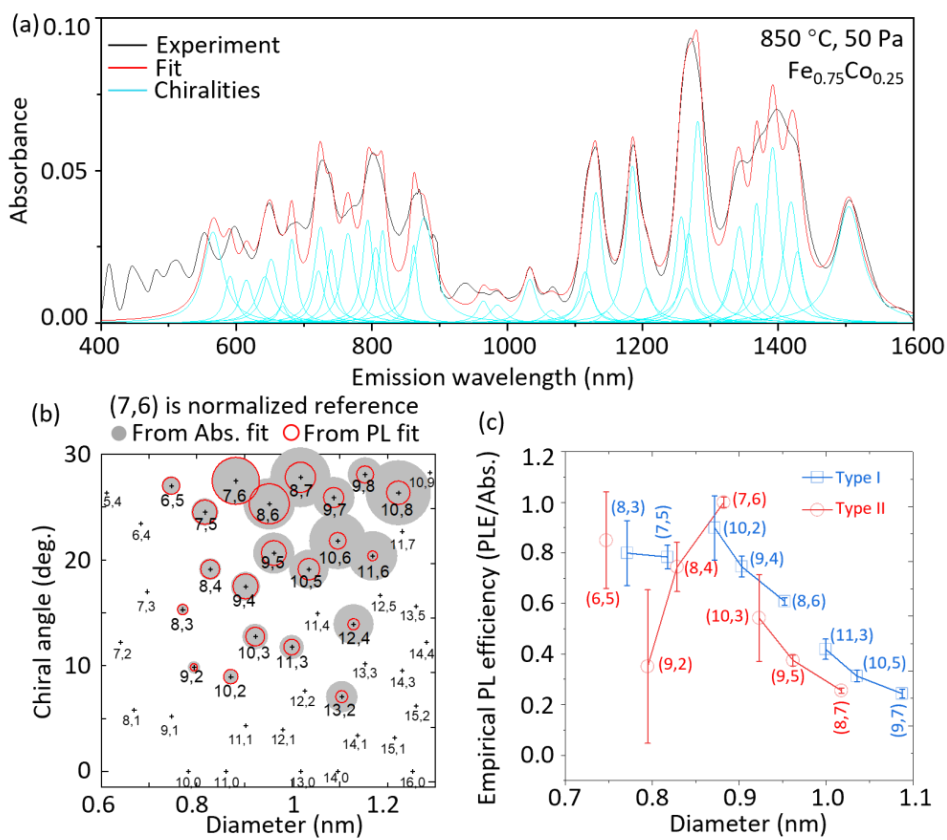

**Figure S19.** (a) Absorption spectrum and corresponding peak deconvolution results for SWCNTs grown at 850 °C and 50 Pa ($Fe_{0.75}Co_{0.25}$ catalyst). The experimental spectrum (black), overall fit (red), and individual chiral contributions (cyan) are shown. (b) 2D chirality distribution map showing the relative intensities of each chirality, with tube diameter on the x-axis and chiral angle on the y-axis. Circle sizes correspond to relative intensities obtained from absorption fitting (gray) and PLE fitting (red outlines), with (7,6) used as a normalization absorption vs. PLE results. (c)

Empirical PL efficiency (PLE/Abs) as a function of diameter, showing reduced PL efficiency and increased fluctuation for large-diameter nanotubes compared to those grown at 600 °C.

Based on the data presented in **Fig. S18** and **S19**, we first verified the reliability of the fitting procedure by comparing the experimental absorption spectra with the fitted results (**Fig. S18a-S19a**). The excellent agreement between the experimental data and the fitted curves demonstrates that the applied deconvolution method accurately captures overlapping peaks and extracts individual chirality intensities. Contributions from metallic SWCNTs (< 600nm) were not included, as they are not present in the PLE maps. Moreover, also a small fraction of larger diameter SWCNTs (>1400 nm in **Fig. S18-S19**) were not included as they were barely visible in the PLE maps and could therefore not be fitted.

Following this validation, we visualized the chirality distribution in a two-dimensional diameter–chiral angle plot (**Fig. S18b and S19b**), where the position of each circle corresponds to the geometric parameters of a specific chirality, and the circle size represents the relative intensity derived from absorption (gray) and PLE fitting (red outline). For the sample grown at 600 °C, the most intense chiral species are concentrated in the diameter range of 0.7–0.95 nm and chiral angles of 10–25°, indicating strong chirality selectivity. In contrast, the sample grown at 850 °C exhibits a broader distribution extending to larger diameters and lower chiral angles, reflecting the influence of increased catalyst particle size and altered catalyst activity at higher temperatures. The close match between the distributions derived from PLE and absorption fitting confirms that the fitting method is capable of faithfully reproducing chirality distributions in geometric parameter space.

Finally, the quantum yields (defined as the ratio of PLE intensity to absorption intensity) for various chiral species were compared, as shown in **Fig. S18-S19c**. At 600 °C, small-diameter chiralities such as (7,6), (8,4), and (6,5) display relatively high PL efficiencies. Under 850 °C growth conditions, the PL efficiencies are significantly lower, and the error bars become larger across multiple chiralities. It is notable that even chiral species with strong PL efficiencies (such as (6,5) and (8,3)) exhibit considerable scatter. Normally, this increased variability can be attributed to the amplified relative uncertainty caused by weaker absorption peaks and poor background subtraction. A comprehensive summary of peak offsets and PL efficiencies for various dispersants and growth conditions is provided in **Table S1**.